%% file: NdBiPt_Mueller.tex
\documentclass[aps,prb,reprint,showpacs,groupedaddress,superscriptaddress]{revtex4-1}
\usepackage{graphicx}
\usepackage{dcolumn}
\usepackage{bm}
\usepackage{amssymb}

\begin{document}

\title{Magnetic structure of the antiferromagnetic half-Heusler compound NdBiPt}
%\input author_list.tex

%\author{}
%\thanks{}
%\affiliation{}
\author{R. A. Müller}
\affiliation{D\'epartement de Physique, Universit\'e de Montr\'al, Montr\'eal, Canada}
\altaffiliation{Regroupement Qu\'eb\'ecois sur les Mat\'eriaux de Pointe (RQMP)}
\author{A. Desilets-Benoit}
\affiliation{D\'epartement de Physique, Universit\'e de Montr\'al, Montr\'eal, Canada}
\altaffiliation{Regroupement Qu\'eb\'ecois sur les Mat\'eriaux de Pointe (RQMP)}
\author{N. Gauthier}
\affiliation{D\'epartement de Physique, Universit\'e de Montr\'al, Montr\'eal, Canada}
\altaffiliation{Regroupement Qu\'eb\'ecois sur les Mat\'eriaux de Pointe (RQMP)}
\altaffiliation[Current Address:]{Laboratory for Scientific Developments and Novel Materials, Paul Paul Scherrer Institut, Villigen, Switzerland}
\author{L. Lapointe}
\affiliation{D\'epartement de Physique, Universit\'e de Montr\'al, Montr\'eal, Canada}
\altaffiliation{Regroupement Qu\'eb\'ecois sur les Mat\'eriaux de Pointe (RQMP)}
\author{A. D. Bianchi}
\affiliation{D\'epartement de Physique, Universit\'e de Montr\'al, Montr\'eal, Canada}
\altaffiliation{Regroupement Qu\'eb\'ecois sur les Mat\'eriaux de Pointe (RQMP)}
\email[Email]{andrea.bianchi@umontreal.ca}
%\homepage[Web]{http://phys.umontreal.ca/repertoire-departement/vue/bianchi-andrea/}
%
\author{T. Maris}
\affiliation{D\'epartement de Chimie, Universit\'e de Montr\'eal, Montr\'eal, Canada}
\author{R. Zahn}
\affiliation{Hochfeld-Magnetlabor Dresden (HLD-EMFL), Helmholtz-Zentrum Dresden-Rossendorf, Dresden, Germany}
\author{R. Beyer}
\affiliation{Hochfeld-Magnetlabor Dresden (HLD-EMFL), Helmholtz-Zentrum Dresden-Rossendorf, Dresden, Germany}
\author{E. Green}
\affiliation{Hochfeld-Magnetlabor Dresden (HLD-EMFL), Helmholtz-Zentrum Dresden-Rossendorf, Dresden, Germany}
\author{J. Wosnitza}
\affiliation{Hochfeld-Magnetlabor Dresden (HLD-EMFL), Helmholtz-Zentrum Dresden-Rossendorf, Dresden, Germany}
\author{Z. Yamani}
\affiliation{Canadian Neutron Beam Centre, National Research Council, Chalk River, Canada}
\author{M. Kenzelmann}
\affiliation{Laboratory for Scientific Developments and Novel Materials, Paul Paul Scherrer Institut, Villigen, Switzerland}

\date{\today}

\begin{abstract}
We present results of single-crystal neutron-diffraction experiments on the rare-earth,
half-Heusler antiferromagnet (AFM) NdBiPt. This compound exhibits an AFM phase transition
at $T_N=2.18$ K with an ordered moment of $1.78(9)$~$\mu_B$ per Nd atom. The magnetic moments
are aligned along the $[001]$ direction, arranged in a type-I AFM structure with ferromagnetic
planes, alternating antiferromagnetically along a propagation vector $\tau$ of $(100)$. The
$R$BiPt ($R$ = Ce - Lu) family of materials has been proposed as candidates of a new family
of antiferromagnetic topological insulators (AFTI) with magnetic space group that corresponds
to a type-II AFM structure where ferromagnetic sheets are stacked along the space diagonal.
The resolved structure makes it unlikely that NdBiPt qualifies as an AFTI.

\end{abstract}

\maketitle

\section{Introduction}
A usual concept in physics is the occurrence of some form of symmetry breaking at phase
transitions between different states of matter.
In 1980, Klaus von Klitzing widened that concept by describing a new quantum state of
matter which does not follow this pattern, but shed light on a new family of materials,
only characterized by their Hilbert-space topology. In this new state of matter, the
bulk of a two-dimensional sample stays insulating, whereas along its edges a unidirectional
current is circulating, giving rise to the quantum  Hall effect in a two-dimensional
electron gas (2DEG). Inspired by the mathematical field of topology, the quantized
conductivity of such a material can be associated with a topological invariant.
In mathematics, such an invariant describes a property of a topology that remains
unchanged under homeomorphisms. For example, the number of holes in a two-dimensional
manifold cannot be changed by stretching it.

In solid-state physics, we can adapt this concept of smooth deformations to the
topology of the Hilbert space, which describes the band structure of an insulator.
As long as these transformations are adiabatic, the topological invariant will not
change, and, therefore, the band gap at the Fermi level of the material remains
unaffected. While the quantum Hall state in a 2DEG requires an applied magnetic
field, in the case of a Hg/CdTe quantum well, strong spin-orbit coupling acts as
an effective field.\cite{Bernevig2006} If the well is thinner than a critical
value $d_c$, it behaves like a conventional insulator. For $d_{QW} > d_c$, the
topological invariant changes and a single pair of helical edge stages that form
a Kramers pair, counter-propagate on the same edge. In consequence, the
magneto-transport in such quantum wells shows steps.\cite{Konig2007}

Spin-orbit coupling is also at the origin of topological insulators in three
dimension.\cite{ Moore2010, Hasan2010, Xiao2010b} Experimentally, spin- and
angle-resolved photoemission spectroscopy on bismuth doped with antimony
showed the presence of metallic surface states, as well as a spin texture.\cite{Hsieh2009c}
At the same time, \emph{ab initio} calculations predict a small gap in the
electronic spectrum for the bulk of this material.\cite{Zhang2009a}

Recently, theorists have brought forward several propositions suggesting, that
half-Heusler compounds, showing antiferromagnetic order, could host a new class
of topological insulators. Mong \textit{et al}.,~\cite{Mong2010} described a new
symmetry class, where both time-reversal and lattice translational symmetry of
an antiferromagnet are broken, yet their product is preserved, resulting in a new
antiferromagnetic topological insulator (AFTI) phase. The broken time-reversal
symmetry of AFTI is what distinguishes them from conventional topological
insulators, where the time-reversal symmetry has to be present for the surface
states to occur, which forbids magnetic order.
Described in their paper as model B,\cite{Mong2010} the orientation of the magnetic
moment can introduce a net magnetization between intermediate non-magnetic sites
creating an Aharonov-Bohm-like flux which acts as Rashba spin-orbit coupling,
resulting in a non-trivial topological phase.

Heusler and the derivative half-Heusler materials can be characterized as semi-metals
displaying insulating or semi-metallic behaviour in electrical transport measurements.
This behaviour agrees with band-structure calculations, which for many of these
compounds show a single band crossing the Fermi surface, which lead to the proposition
that conventional topological insulators can be found in this class of compounds.\cite{Chadov2010,Al-Sawai2010,Xiao2010b,Li2011} \textit{R}BiPt materials,
where \textit{R} is a rare earth, first reported in detail in 1991,\cite{Canfield1991}
display a whole set of emergent behaviours ranging from a massive electron state in
YbBiPt,\cite{Fisk1991} to superconductivity without inversion symmetry in
LaBiPt,\cite{Goll2002} LuBiPt,\cite{Tafti2013}, and YBiPt,\cite{Butch2011,Bay2012,Bay2014}
to CeBiPt which shows a magnetic field induced change of the Fermi surface.\cite{Kozlova2005}
This also prompted investigations of the \textit{R}BiPd~\cite{Gofryk2007} versions
which led to the discovery of superconducting LuBiPd, which shows an anomaly in
the electronic specific heat at the superconducting transition, and weak anti-localization
in the magnetic-field dependence of the electrical resistivity, which is characteristic
for 2D conduction.\cite{Xu2014}

Angle-resolved photoemission experiments (ARPES) on Lu, Dy, and GdBiPt have shown
indications of metallic surface states that differ from the bulk band structure.
Liu \textit{et al}.,~\cite{Liu2011} found that within their resolution an even
number of bands cross at the chemical potential, making surface states vulnerable
to non-magnetic backscattering and these materials should, therefore, not be
qualified as strong topological insulators.
An inelastic x-ray\cite{Kreyssig2011a} as well as a powder neutron diffraction
experiment\cite{Muller2014} on GdBiPt indicate a doubling of the unit cell along
its space diagonal with the moments arranged in ferromagnetic sheets,\cite{Muller2014}
normal to the $[111]$ direction, leading to a path asymmetry for %plane
hopping between
non-magnetic sites, as proposed by Mong \textit{et al}., and, therefore, making this
material a strong candidate for the AFTI phase.

This has prompted us to carry out single-crystal neutron and X-ray diffraction,
as well as thermodynamic and transport experiments, to determine the magnetic
structure of NdBiPt, as its crystalline structure has all the necessary
symmetries for being an AFTI.

\section{Samples and Experiment}
NdBiPt was grown using Bi flux. Nd, Bi, and Pt of high purity were placed in a
ceramic crucible in the ratio 1:15:1 which was then sealed in a quartz ampoule
under argon atmosphere. The melt was kept at $1200\,^{\circ}\mathrm{C}$ for two
days and then cooled down to $550\,^{\circ}\mathrm{C}$  over a week, after which
the ampoules were taken out of the furnace and centrifuged to separate the flux
from the crystals.

Magnetic measurements were taken between 1.8 and 300~K in an applied field of
0.1~T using a Quantum Design VSM SQUID magnetometer. Resistivity was measured
in the same temperature range with a Quantum Design PPMS using four-point
contacts. The specific heat $C_p$ was measured in a $^3$He insert PPMS
using a standard puck but purpose-built electronics.

Single-crystal X-ray diffraction data were collected at 150~K on a Bruker D8
VENTURE diffractometer with a CMOS PHOTON 100 detector and a liquid-metal jet
x-ray source using Ga radiation ($\lambda=1.3414$~\AA).
The data set was collected using a combination of $\omega$ and $\phi$ scans
with a step size of $1^{\circ}$, and 1~s exposure per frame. Data collection
and unit-cell lattice parameters determination were performed with the
\textsc{apex2} suite.\cite{APEX} Final lattice-parameter values and integrated
intensities were obtained using \textsc{saint} software, and a multi-scan
absorption correction was applied with \textsc{sadabs}.\cite{Sheldrick2008a}
The structure was refined with \textsc{shelxl} version 2014/3.\cite{Sheldrick2008}

For the single-crystal neutron-diffraction experiment we co-aligned three
crystals of the size of the order $2 \times 1 \times 1$~mm$^3$  on an aluminum
plate. We oriented our crystals to be able to scan the $(hhl)$ scattering plane
given the extinction rules of the NdBiPt crystalline structure. Also, this
scattering geometry allows us to distinguish between the type-I AFM order,
seen in the isostructural CeBiPt \cite{Wosnitza2006} and  type-II AFM order,
as proposed by Mong \textit{et al}., in Ref.~[\onlinecite{Mong2010}] and
observed in GdBiPt.\cite{Muller2014} The experiment was carried out on the
C5 triple-axis spectrometer at the Canadian Neutron Beam Centre in Chalk River.
A vertically focusing pyrolytic graphite (PG) $(002)$ monochromator and a flat
PG$(002)$ analyzer crystal were used with a fixed final neutron energy of
$E_f = 14.56$~meV, with no collimation, and collimations of $0.8^{\circ}$,
$0.85^{\circ}$, and $2.4^{\circ}$. Two PG filters were placed in the diffracted
beam after the sample to eliminate higher-order wavelength contamination of
the beam. The sample was sealed under helium gas in an aluminium can and mounted
in a close-cycle $^3$He heliox displex cryostat that allowed cooling the sample
down to 0.3~K.

\section{Results and Analysis}
\subsection{X-ray diffraction}%
NdBiPt crystallizes in the cubic half-Heusler crystal structure with the space
group $F\bar{4}3m$. \cite{Canfield1991} This structure consists of four
interpenetrating \textit{fcc} lattices shifted by
$[\frac{1}{4},\frac{1}{4},\frac{1}{4}]$, where the $[\frac{1}{2},\frac{1}{2},\frac{1}{2}]$
position is an ordered vacancy. The refinement of our single-crystal x-ray patterns
confirms this structure (for details see Tables I, and II in the SOM). The compound
has a lattice constant of 6.7613(2)~\AA\  with the Nd$^{3+}$ ion located on the
$[\frac{1}{4},\frac{1}{4},\frac{1}{4}]$ (4c), Bi on the
$[\frac{3}{4},\frac{3}{4},\frac{3}{4}]$ (4d), and Pt on the $[0,0,0]$ (4a)
position, and permutations of $[0,\frac{1}{2},\frac{1}{2}]$ (corresponding to
the column D of Table II of the SOM).

In a non-centrosymmetric structure, anomalous x-ray scattering leads to different
intensities for so-called \emph{Friedel} pairs, such as $(hkl)$ and
$(\bar{h}\bar{k}\bar{l})$. The refinement confirms the original structure
(see Fig.~1 of the SOM), resulting in an $R1$ value of 0.0582, where $R1$ is
the difference between the experimental observed squares of the structure
factors for all observed peaks, and the respective calculated values. Also
the Flack parameter for the original structure is 0.28(3), which is the
absolute structure factor. This is in contrast to a $R1$ value of 0.0800
and Flack parameter of 0.72(4) for the inverted structure, as listed in
Table II of the SOM. Please note that a Flack parameter should be close to
0 for a correct structure and close to 1 for an inverted structure.

\subsection{Magnetic and Transport Properties}
\begin{figure}
 \includegraphics[width=0.45\textwidth]{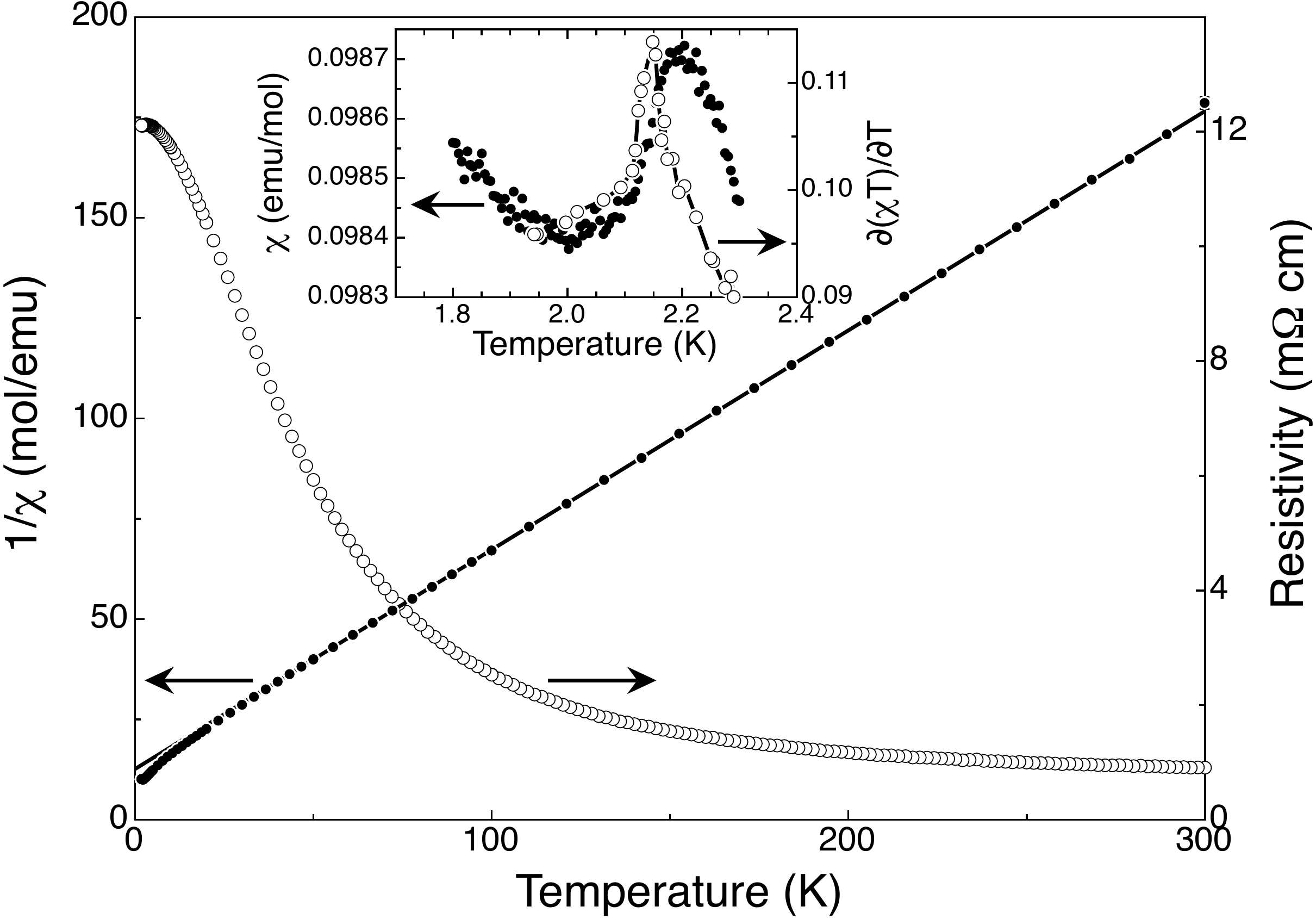}
  \caption{Inverse magnetic susceptibility measured at 0.1~T and resistivity (at 0~T) as a function of temperature. The inverse susceptibility has been fitted with a Curie-Weiss law in the high-temperature regime yielding $\Theta_{\mathrm{W}}=-23$~K with an effective moment of $\mu_{\mathrm{eff}}=3.73$~$\mu_{\mathrm{B}}$ Inset: Temperature derivative $\partial (T \chi)/ \partial T$ of the magnetic susceptibility showing a sharp peak at the critical temperature $T_\mathrm{N}$ of 2.18~K.}
  \label{fig:transport}
\end{figure}

NdBiPt is a semi-metal with a very low charge-carrier density, and a high
charge-carrier mobility.\cite{Morelli1996} For the temperature range 50 to 300~K,
the magnetic susceptibility $\chi=\frac{M}{H}$ measured in an applied field of 0.1~T
shows a Curie-Weiss behaviour with a Curie-Weiss temperature $\Theta_{\mathrm{W}}$
of -23~K (see Fig.~\ref{fig:transport}), and an effective magnetic moment
$\mu_{\mathrm{eff}}$ of 3.73~$\mu_{\mathrm{B}}$ consistent with the theoretical
value of 3.62~$\mu_{\mathrm{B}}$ for a free Nd$^{3+}$ ion. The inset of
Fig.~\ref{fig:transport} shows  $\chi(T)$ in the temperature range between
1.8 and 2.4~K, where the main features are a maximum at 2.2~K and a subsequent
point of inflection at 2.18~K, confirming  antiferromagnetic order  with a
N\'eel temperature $T_\mathrm{N}$ of  $2.18$~K.\cite{Fisher1968}
All three measurements: Specific heat $C_p$ (see Fig.~\ref{fig:CP}), electrical
resistivity $\partial \rho/\partial T$ (not shown), as well as the magnetic
susceptibility $\partial (T \chi)/ \partial T$ (inset of Fig.~\ref{fig:transport})
show a discontinuity at the same critical temperature $T_{\mathrm{N}}$, giving
evidence for the high quality of our samples.

\subsection{Neutron diffraction}%
\label{sec:neutron}
\begin{figure}
\includegraphics[width=0.45\textwidth]{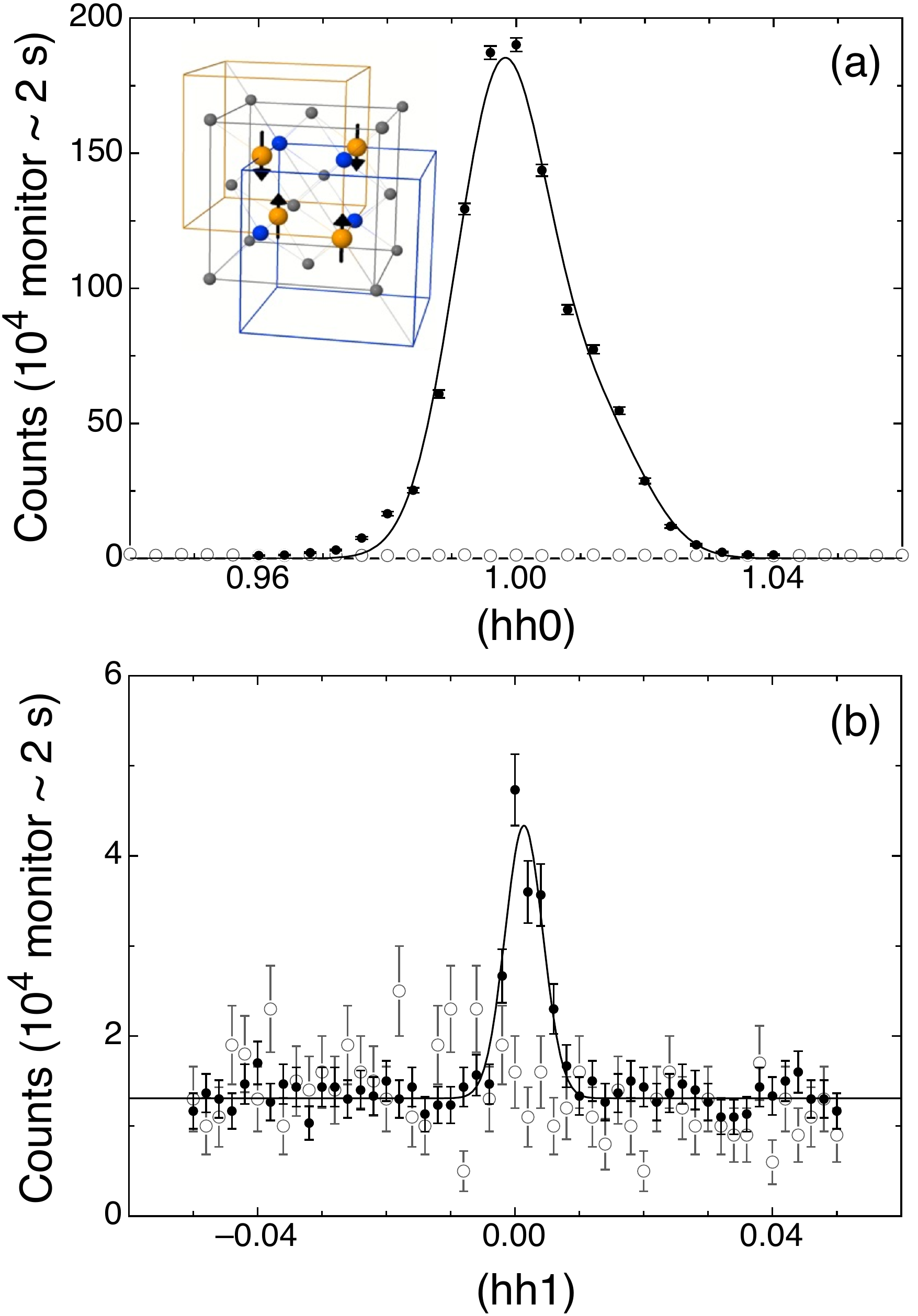}
  \caption{ (Color online) (a) Scan along the $(hh0)$ direction at 0.3~K showing the $(110)$ magnetic peak (full circles). The open circles are the signal  at 5~K above $T_\mathrm{N}$ (open circles). The inset shows the crystal structure of NdBiPt including the three sub-lattices for the three different atomic species. (b) Magnetic signal at 0.3~K (full circles) below $T_{\mathbf{N}}$ due to secondary scattering of neutrons which were first diffracted by the $(111)$ nuclear peak (signal at 5~K shown as open circles). This position in reciprocal space corresponds to a $(001)$ magnetic  peak, however, with a much reduced intensity. }
  \label{fig:direc}
\end{figure}

Neutron-diffraction data was collected between  0.3  and 5~K. We used a linear
fit for the  background. Our measurements show slight mosaic due to a small
misalignment of the three crystals of about one degree, as can be seen in the
peak shape in Fig.~\ref{fig:direc}(a). To correct for the mosaic the peaks were
fitted with a double Gaussian:
\begin{eqnarray}
G(x) & = & B+A\cdot e^{ \frac{-4\cdot \ln{2}\cdot |x-x_0|^2}{s^2}}\nonumber\\
& &\times\bigg\{ 1+\frac{1}{R}e^{ \frac{4\cdot \ln{2}\cdot(\Delta^2+2|x-x_0|\cdot \Delta)}{s^2}}\bigg\},
\label{eq:zero}
\end{eqnarray}
where $B$ corrects for an imperfect background subtraction. $A$ is the amplitude
and $x_0$ denotes the center position of the dominant peak. The parameter $s$
represents the full width at half maximum (FWHM), $R$ is the ratio in intensity
of the two peaks and $\Delta$ represents the distance between the two peak centers
along $x$.

\begin{figure}
\includegraphics[width=0.45\textwidth]{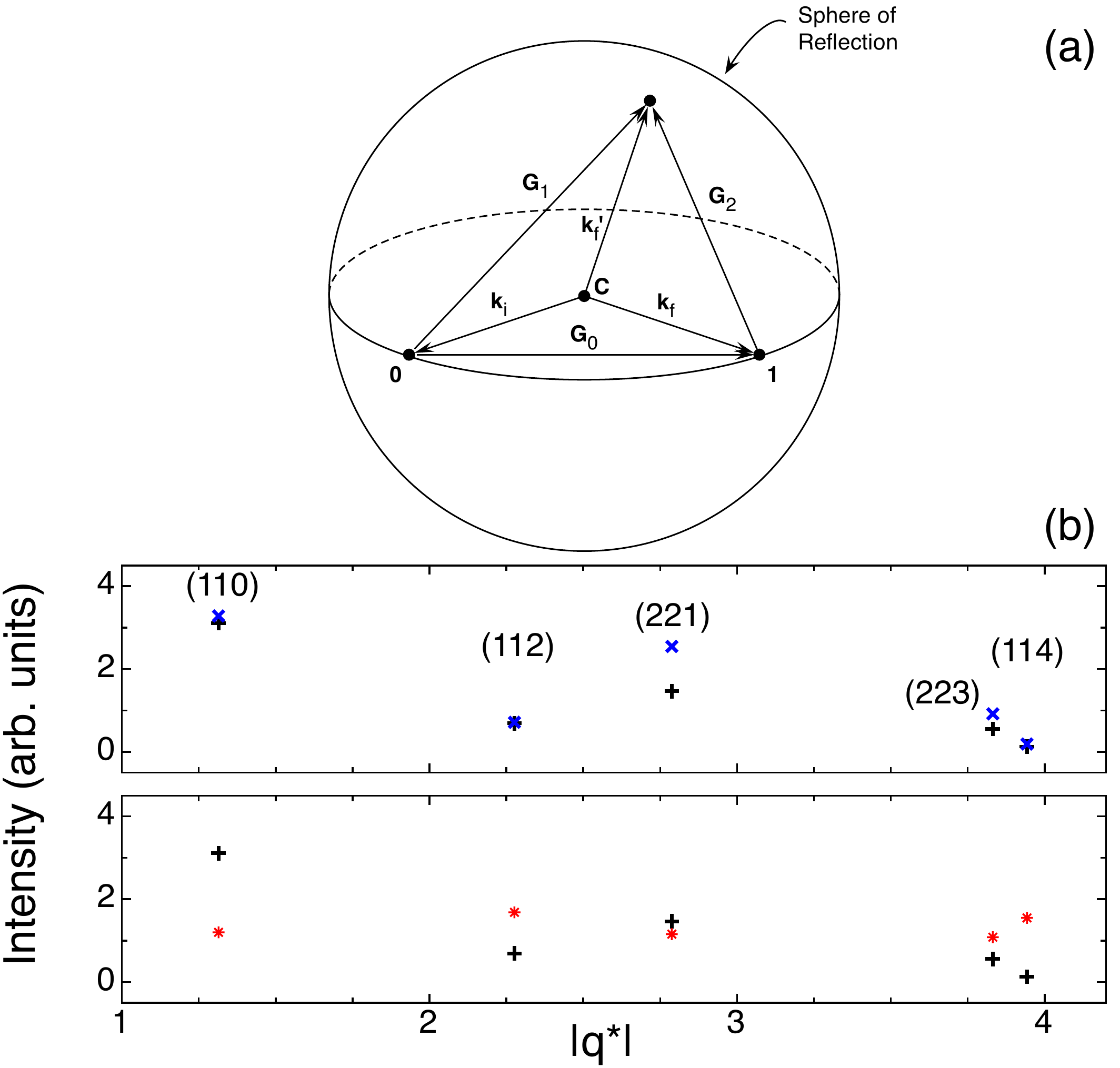}
  \caption{ (Color online) (a) Neutrons diffracted by $\bold{G}_1$
    undergo a second scattering by a reciprocal lattice vector
    $\bold{G}_2=\bold{G}_0 -\bold{G}_1$.\cite{Shirane2002} (b) The
    measured intensities are shown as crosses circles. The diagonal
    crosses reflect the refined intensities using \textsc{FullProf}
    with the correct basis  (top), and the stars with the wrong basis
    set (bottom).For a propagation vector $\tau$ of  $(100)$ the 
    correct basis corresponds to magnetic moments which are aligned along the crystallographic $c$-axis.}
 \label{fig:meas_vs_calc}
\end{figure}

All the observed magnetic peaks could be indexed as integer fractions of the nuclear
peaks which is evidence for a commensurate magnetic structure
[see Fig.~\ref{fig:meas_vs_calc}(b)].  For spins located on an \textit{fcc}
lattice, only four types of commensurate antiferromagnetic order are
possible.\cite{Yildirim1998}  To determine the direction of the magnetic
moment, we compare the intensities of the $(110)$ peak with those of the
$(001)$ peak. The intensities observed at these two Bragg spots indicate
that the magnetic moment is aligned parallel to the momentum of the incoming
neutron beam, along the $[001]$-direction, as shown in the inset of
Fig.~\ref{fig:direc}(a).

Due to the cubic structure of the crystal, the magnetic moment can point along
any of the six edges of the cube, giving rise to three equally probable magnetic
domains, which are equivalent by symmetry. In our scattering geometry, the
structure factor is such that the signal from for two of these domains are canceled,
leaving only the $[001]$-domain observable. From this, we conclude that the magnetic
moment of the Nd$^{3+}$ ion points normal to the $\{100\}$ family of planes. This
means that in NdBiPt the moment lies along the directions $[100]$, or the equivalent
$[010]$, and $[001]$ directions [inset of Fig.~\ref{fig:direc}(a)]. As we have
not reason to assume that one of these domains is preferentially populated, such
as can be achieved through the application of mechanical strain to the sample, or
by applying a magnetic field, we expect all three domains have the same probability
to occur. We accounted for the existence of domains when we calculated the size of
the ordered magnetic moment. We also would like to point out, that these domains
are large, as the width of the magnetic peaks is comparable to the width of the
nuclear peaks which are limited by the instrument and the particular instrument
set-up we used. In principle, a so-called multi-$\vec{k}$-structure with multiple
propagation vectors could also explain the observed peak intensities, however, we
believe this to be unlikely due to the most probable Heisenberg nature of the
magnetic interactions in NdBiPt.\cite{Rossat-Mignod1987}

The observed structure has ferromagnetic ordered planes with alternating spin
direction along the propagation vector $\tau=(100)$, similar to what previously
has been observed in the isostructural compound CeBiPt.\cite{Wosnitza2006} However,
the magnetic order required for the AFTI phase has to have a magnetic-moment
component that lies in the Nd-plane of the structure, as this would add a net
magnetic field which has to be accounted for in the spin-orbit Hamiltonian with
an additional Aharonov-Bohm phase that is proportional to the in-plane
magnetization.\cite{Mong2010} We find that the moments in NdBiPt are aligned
perpendicular to the Nd layer, resulting in a zero net in-plane magnetization,
and, therefore, the magnetic order has no impact on the strength of the spin-orbit
interaction, as the spins on two neighbouring Nd-atoms always cancel each other.
We, therefore, conclude that NdBiPt does not qualify as representative of the
$S$-symmetry class as described in the article of Mong \emph{et al}.\cite{Mong2010}

\begin{table}
\caption{\label{tab:ireps}Real (BASR) and imaginary (BASI) components
  of the basis vectors for the two permitted commensurable
magnetic structures obtained from \textsc{BasIreps} and the resulting RF-factors
from the \textsc{FullProf} refinment, for the space group
$F\overline{4}3m$ with an ordering wave vector $\tau$ of
$[001]$, and Nd$^{3+}$ occupying
the $4c$ crystallographic site (see SOM). }
\begin{ruledtabular}
\begin{tabular}{lccccc}
& Set 1& RF-factor& \multicolumn{2}{c}{Set 2}&RF-factor\\[1ex]
\hline
& & & & &\\
BASR  &   (0 0 1) &11.5  &  (1 0 0) & (0 1 0)&47.2\\[1ex]
BASI & (0 0 0) & & (0 0 0) & (0 0 0)&\\[1ex]
\end{tabular}
\end{ruledtabular}
\end{table}

We performed a single-crystal refinement of the integrated peak intensities using
the \textsc{FullProf} suite.\cite{Rodriguez-carvajal1993} A representational
analysis using \textsc{BasIreps} for the space group $F\bar{4}3m$ with a propagation
vector $\tau$ of $(001)$ of this type-I AFM structure, i.e., the decomposition of
the magnetic representation in terms of non-zero irreducible representations of all
the symmetry groups that leave $\tau$ invariant into the so-called \emph{little groups}.
This analysis results in two sets of basis functions which are listed in
Table \ref{tab:ireps}.
%
%
%\begin{equation}
%\bold{S}=C \cdot [\bold{BasR} +i  \bold{BasI}]
%\label{eq:bas}
%\end{equation}
%The two basis functions of set 2 represents the two inverted structures
%possible. Due to the fact that we used powder these are
%indistinguishable in the refinement and we are left with a single
%parameter $C$ as refinable quantity.
%
%

The refinement of nuclear peaks followed by the magnetic refinement results in a
magnetic moment of $1.78(8)$~$\mu_{\mathrm B}$ with an RF factor of 11.5, where
the RF factor is the difference between the observed structure factors and the
square root of the calculated structure factors. The difference between the two
representations is illustrated in Fig.~\ref{fig:meas_vs_calc}(b). It can be seen,
that the $(221)$ peak shows a higher intensity than the $(112)$ peak in agreement
with the experimentally observed intensities, as expected for the magnetic structure
presented in the inset of Fig.~\ref{fig:direc}(a).   The value for the magnetic moment
which we obtain from our refinement is considerably lower than the value of
3.8~$\mu_{\mathrm{B}}$ obtained from Curie-Weiss analysis of the high-temperature
susceptibility data. This reduction can be accounted for by crystalline electric
field effects (CEF, see Sec.~\ref{sec:CEF}).

We did observe a small magnetic signal at the $(001)$ position below the critical
temperature, as shown in Fig.~\ref{fig:direc}(b). We can exclude higher harmonics
of the fundamental wavelength as the source of this signal due to the presence of
PG filters. This led us to the conclusion that the observed intensity must result
from second scattering: The incoming beam is first diffracted by the nuclear $[111]$
plane, as schematically shown in Fig.~\ref{fig:meas_vs_calc}(a). The diffracted beam
does now allow for a small magnetic intensity at the same position, which would
correspond to a $(001)$ magnetic reflection of the primary beam.

An estimate of the strength of a $(001)$ magnetic peak due to secondary scattering
can be obtained by using the outgoing flux from the $(111)$ nuclear peak, as the
incident beam that causes the $(001)$ reflection. This estimate results in an
integrated intensity,
%$I_{(001)}^{\mathrm{calc}}=1.1\cdot~ I_{(001)}^{\mathrm{obs}}$%,
which is only about 10\% higher than the observed one, thus substantiating our
conjecture.

\begin{figure}
\includegraphics[width=0.45\textwidth]{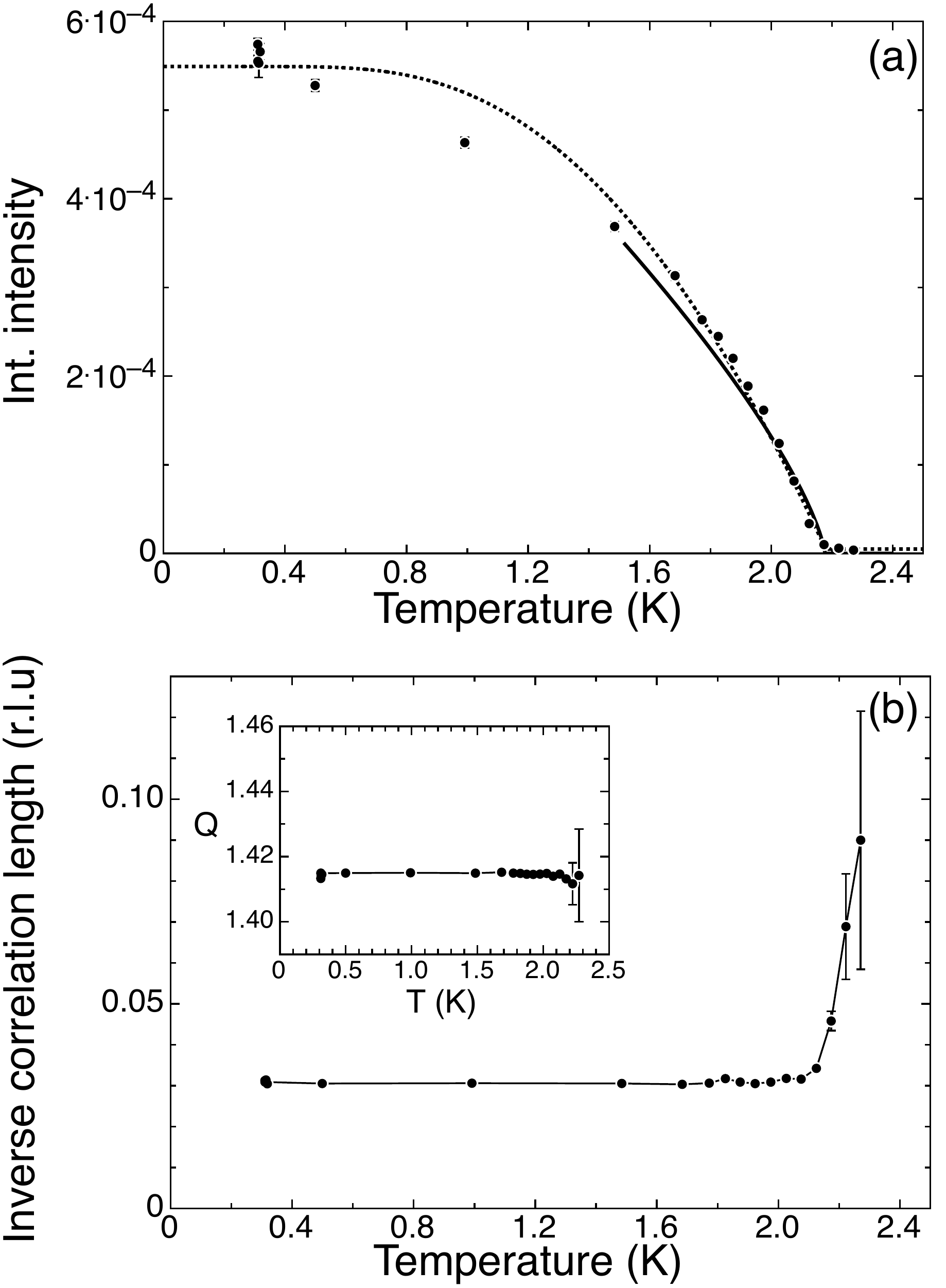}
\caption{ (a) Temperature dependence of the integrated intensity of the $(110)$ magnetic Bragg reflection. The solid line shows the scaling-law fit according to  Eq.~(\ref{eq:one}) used to determine $T_{\mathrm{N}}$. The dashed line is fit of the intensity to the Brillouin function of the CEF doublet. (b) Temperature dependence of the inverse correlation length. Inset: Peak position in $q$-space.  The solid lines are guides to the eye.}
\label{fig:scaling}
\end{figure}

Fig.~\ref{fig:scaling}(a) shows the temperature dependence of the integrated
intensity of the $(110)$ magnetic peak as we cross the transition temperature.
To obtain the N\'eel temperature of $T_{\mathbf{N}}=2.177 \pm 0.005 $~K, the
data was fitted to the scaling law in the temperature range between 1.6 and 2.3 K
[see Fig.~\ref{fig:scaling}(a)]: %\cite{kadanoff}:
\begin{eqnarray}
I&=&C \bigg(1-\frac{T}{T_\mathrm{N}}\bigg)^{2\beta}\quad,
\label{eq:one}
\end{eqnarray}
yielding a critical exponent of $\beta=0.370 \pm 0.003$, which is close to the
value of $\beta=0.369(2)$ expected for a three dimensional Heisenberg antiferromagnet.\cite{Pelissetto2002}
Figure~\ref{fig:scaling}(b) shows the temperature dependence of the Gaussian
peak width along the $(110)$ direction, which is proportional to the average
inverse correlation length $1/\xi$. One can see that $\xi$ diverges, as we
cross the transition temperature, indicating long-range magnetic order.

\section{Crystalline Electric Field Effects}
\label{sec:CEF}
\begin{figure}
\includegraphics[width=0.45\textwidth]{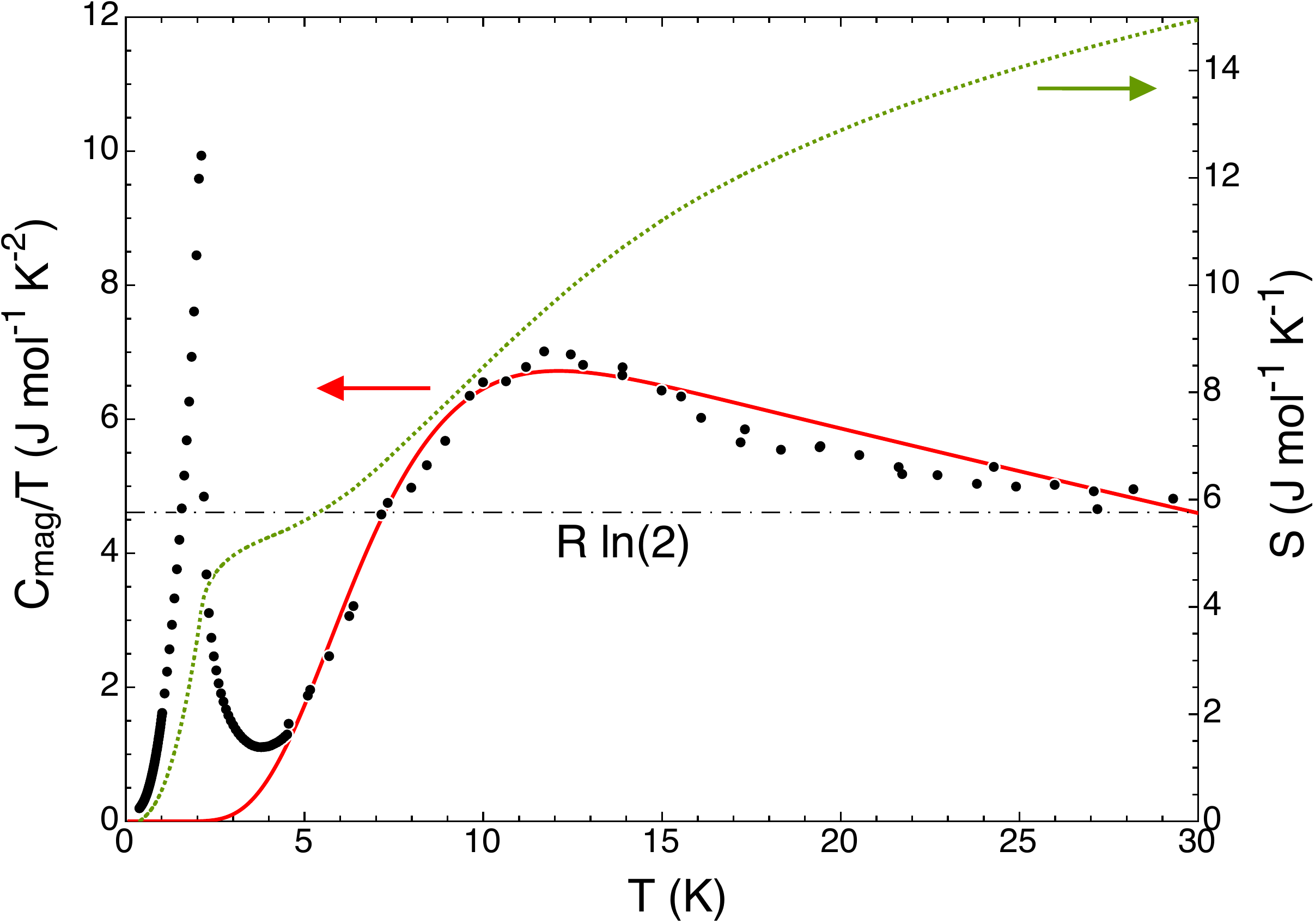}
\caption{(Color online) Magnetic contribution to the specific heat $C_{\mathrm{mag}}$ shown as $\frac{C_{\mathrm{mag}}}{T}$~vs.~$T$. The solid line is the best fit of a Schottky anomaly by using all possible energy eigenvalue configurations obtained by solving the CEF Hamiltonian. The dotted line shows the temperature dependence of the magnetic entropy $S_{\mathrm{mag}}$, which displays a plateau at $R \ln{2}$ indicating the $\Gamma_6$ doublet as the CEF ground state.}
\label{fig:CP}
\end{figure}
As noted in Sec.~\ref{sec:neutron}, the ordered magnetic moment observed in neutron
scattering of $1.78(9)$~$\mu_{\mathrm B}$, is strongly reduced  compared to the
free-ion value of Nd$^{3+}$ of 3.62~$\mu_{\mathrm{B}}$. Since in our scattering
geometry only the signal from one of the three domains contributes, we effectively
only observed $1/3$ of the total magnetic moment in our measurement. Here we assumed
that all three domains have an equal probability, as the observed magnetic structure
does not brake any additional symmetry, besides doubling the unit cell. Such a
reduction of the magnetic moment is often observed in intermetallic compounds due
to crystalline electric field effects (CEF). As similar moment reduction to CEF
effects was reported for CeBiPt, where the ordered moment corresponds to the
magnetic moment of the  $\Gamma_8$ ground state of the
Ce$^{3+}$-ion.\cite{Wosnitza2006,Goll2007}

The effect of the CEF is to lift the 10-fold degeneracy of the $J=\frac{9}{2}$
multiplet of the Nd$^{3+}$-ion through an electrostatic interaction. For a
Nd$^{3+}$-ion sitting in a cubic environment, the CEF splitting is expected
to result in a new ground state consisting of a $\Gamma_6$ doublet, and two
quartets, $\Gamma_8^{(1)}$ and $\Gamma_8^{(2)}$.\cite{Lea1962}

To further investigate the conjecture that this reduction might be due to CEF,
we carried out specific-heat measurements in zero field from 0.3 to 30~K. The
total specific heat, $C_p =  C_{\mathrm{el}} + C_{\mathrm{ph}} + C_{\mathrm{mag}}$,
is the sum of the electronic contribution $C_{\mathrm{el}} = \gamma T$, the
phonon contribution $C_{\mathrm{ph}}$, and the magnetic contribution
$C_{\mathrm{mag}}$ we are interested in. Due to the large phonon
$C_{\mathrm{ph}}$ and magnetic $C_{\mathrm{mag}}$ contributions in the measured
temperature range, we were not able to determine the electronic contribution
$C_{\mathrm{el}}$ and could only establish that it is below 1~mJ\,mol$^{-1}$\,K$^{-2}$.
Such a low value for $C_{\mathrm{el}}$ is expected due to the low charge-carrier
concentration in NdBiPt. Over the measured temperature range, $C_{\mathrm{ph}}$
can be described by the Debye function:
\begin{eqnarray}
C_{\mathrm{ph}} &=& 9   R \bigg( \frac{T}{\theta_{\mathrm{D}}} \bigg)^3 \int_0^{\frac{\theta_{\mathrm{D}}}{T}} \frac{x^4 e^x}{(e^x-1)^2} dx \quad,
\label{eq:Debye}
\end{eqnarray}
where $R$ is the universal gas constant und $\theta_\mathrm{D}$ the Debye temperature.
As can be seen in Fig.~\ref{fig:CP}, we observe a rather broad magnetic peak between
4.5 and 20~K, which makes it difficult to fit the phonon contribution. We instead chose
to use the $\theta_\mathrm{D}$ value of 122.3~K obtained from a fit of
Eq.~(\ref{eq:Debye}) to the specific-heat data of GdBiPt, which does not have CEF
splitting.\cite{Muller2014} We then scaled the Debye temperature with the square
root of the inverse mass ratio between Gd and Nd. This yields a Debye temperature
of $\theta_\mathrm{D}=123.7$~K for NdBiPt. Fig.~\ref{fig:CP} shows the magnetic
contribution $C_{\mathrm{mag}}= C_p - C_{\mathrm{ph}}$  to the specific heat after
subtraction of the phonon contribution. By integrating the magnetic specific heat
$C_{\mathrm{mag}}$, we can obtain the magnetic entropy $S_{\mathrm{mag}}=\int_0^T
\frac{C_{\mathrm{mag}}}{T}dT$ associated with the CEF ground state which orders.
$S_{\mathrm{mag}}$ shows a plateau at about $R \ln{2}$ corresponding to a
doublet ground state.

To analyze the Stark splitting of our degenerate ground state due to the
crystalline electric field (CEF) we search for solutions of the perturbation
Hamiltonian for an eightfold cubic symmetry. For  $f$-electron configurations
terms up to the sixth order are sufficient:\cite{Lea1962}
\begin{eqnarray}
\mathcal{H}_{\mathrm{CEF}}& =& B_4\big(O_4^{0}+5 O_4^4\big)+B_6\big(O_6^{0}+21  O_6^4\big)\quad.
\label{eq:CEF}
\end{eqnarray}
Here, the $O_n^m$ are the Stevens's equivalent operators and the $B_n$ are the
CEF amplitudes describing the admixture between the different states
$\vert \pm \frac{9}{2} \rangle$,~$\ldots$,~$\vert \pm \frac{1}{2} \rangle$
of the $J$ multiplet.

To determine the ratio between fourth and sixth order terms, we follow the
procedure laid out in Ref.~[\onlinecite{Lea1962}], and substitute
$O_4=O_4^0+5O_4^4$ and  $O_6=O_6^0-21 O_6^4$. Thus, we can rewrite
Eq.~(\ref{eq:CEF}) as:
\begin{eqnarray}
\mathcal{H}_{\mathrm{CEF}}&  = &W\bigg[x\frac{O_4}{F(4)}+(1-|x|)\frac{O_6}{F(6)}\bigg]\quad,
\label{eq:CEF2}
\end{eqnarray}
where $B_4F(4)=Wx$, and $B_6F(6)=W(1-|x|)$ for $-1<x<+1$. This allows us to
fit to the magnetic part of the specific heat $S_{\mathrm{mag}}$ for different
values of $x$ and $W$  (see Fig.~2 in the SOM) in terms of a Schottky anomaly:
\begin{eqnarray}
C_{\mathrm{CEF}}&=&\frac{R}{T^2}\Bigg[\frac{4 \Delta_1^2 e^{-\frac{\Delta_1}{T}}+ 4\Delta_2^2 e^{-\frac{\Delta_2}{T}}}{2+ 4e^{-\frac{\Delta_1}{T}}+4e^{-\frac{\Delta_2}{T}}}\nonumber\\
& & -\bigg( \frac{4 \Delta_1 e^{-\frac{\Delta_1}{T}}+4 \Delta_2 e^{-\frac{\Delta_2}{T}}}{2+4e^{-\frac{\Delta_1}{T}}+4e^{-\frac{\Delta_2}{T}}} \bigg)^2  \Bigg].
\label{eq:schottky}
\end{eqnarray}
For Nd$^{3+}$ with a $J=9/2$, the 10-fold degenerate ground state is lifted into
a doublet $\Gamma_6$ as the ground state  and the two quadruplets $\Gamma_8^{(1)}$
and $\Gamma_8^{(2)}$, which are separated by an energy gap of $\Delta_1$ and
$\Delta_2$, respectively. We obtain a best fit shown as the solid line in
Fig.~\ref{fig:CP} for  $\Delta_1=29$~K and $\Delta_2=72$~K. This allows for two
solutions, one with $x=-0.9650$ and $W/k_B=1.14$~K, and the other for $x=0.140$
and $W/k_B=0.774$~K.

Knowing the values of $x$ and $W$ allows us to calculate the expected
magnetic moment of the $\Gamma_6$ doublet.  For both solutions, this
calculation yields a theoretical value of 1.83~$\mu_{\mathrm{B}}$ for
the ordered moment, which is close to the $1.78(9)$~$\mu_{\mathrm B}$
obtained from neutron diffraction.

\section{Conlusions}
We determined the magnetic structure of the semi-metal NdBiPt, which crystallizes
in a half-Heusler structure. Below the N\'eel temperature $T_{\mathrm{N}}$ of
2.18~K we find an up-down structure of ferromagnetically aligned planes, in which
the spin of the Nd points along the $[001]$ direction, that alternate along the
propagation vector $\tau=(100)$. This type-I structure is common for crystals
belonging to the space group $F\bar{4}3m$. This opens the question why in
GdBiPt,\cite{Kreyssig2011a, Muller2014} YbBiPt,\cite{Ueland2014} and
vanadium-doped CuMnSb \cite{Forster1968,Halder2011} the propagation vector
of the antiferromagnetic structure (AFM) points along $[111]$. However,
the magnetic structure we found in NdBiPt excludes this material from
being a candidate  for the proposed new class of antiferromagnetic
topological insulators  (AFTI).\cite{Mong2010} In NdBiPt, the ground
state is the $\Gamma_6$ CEF doublet which orders, and we find an ordered
moment of $1.78(9)$~$\mu_{\mathrm B}$.
\begin{acknowledgments}
 \input acknowledgement.tex   % input acknowledgement
\end{acknowledgments}
\bibliography{NdBiPt}% Produces the bibliography via BibTeX.
%\appendix*
%
\end{document}

%% file: acknowledgement.tex
% acknowledgement.tex                            9 February 2012 
%
We thank O. Stockert from the MPI-CPFS in Dresden, Germany, for the
disscussion on double scattering. We also acknowledge the help from
Oksana Zaharko from the PSI in Villigen, Switzerland, for her help
with the \textsc{FullProf} refinement. The research at UdeM received support
from the Natural Sciences and Engineering Research Council of Canada
(Canada), Fonds Qu\'eb\'ecois de la Recherche sur la Nature et les
Technologies (Qu\'ebec), and the Canada Research Chair Foundation.
Part of this work was supported by HLD at HZDR, a member of the
European Magnetic Field Laboratory (EMFL).

%% file: NdBiPt_Mueller.bbl
\begin{thebibliography}{40}%
\makeatletter
\providecommand \@ifxundefined [1]{%
 \@ifx{#1\undefined}
}%
\providecommand \@ifnum [1]{%
 \ifnum #1\expandafter \@firstoftwo
 \else \expandafter \@secondoftwo
 \fi
}%
\providecommand \@ifx [1]{%
 \ifx #1\expandafter \@firstoftwo
 \else \expandafter \@secondoftwo
 \fi
}%
\providecommand \natexlab [1]{#1}%
\providecommand \enquote  [1]{``#1''}%
\providecommand \bibnamefont  [1]{#1}%
\providecommand \bibfnamefont [1]{#1}%
\providecommand \citenamefont [1]{#1}%
\providecommand \href@noop [0]{\@secondoftwo}%
\providecommand \href [0]{\begingroup \@sanitize@url \@href}%
\providecommand \@href[1]{\@@startlink{#1}\@@href}%
\providecommand \@@href[1]{\endgroup#1\@@endlink}%
\providecommand \@sanitize@url [0]{\catcode `\\12\catcode `\$12\catcode
  `\&12\catcode `\#12\catcode `\^12\catcode `\_12\catcode `\%12\relax}%
\providecommand \@@startlink[1]{}%
\providecommand \@@endlink[0]{}%
\providecommand \url  [0]{\begingroup\@sanitize@url \@url }%
\providecommand \@url [1]{\endgroup\@href {#1}{\urlprefix }}%
\providecommand \urlprefix  [0]{URL }%
\providecommand \Eprint [0]{\href }%
\providecommand \doibase [0]{http://dx.doi.org/}%
\providecommand \selectlanguage [0]{\@gobble}%
\providecommand \bibinfo  [0]{\@secondoftwo}%
\providecommand \bibfield  [0]{\@secondoftwo}%
\providecommand \translation [1]{[#1]}%
\providecommand \BibitemOpen [0]{}%
\providecommand \bibitemStop [0]{}%
\providecommand \bibitemNoStop [0]{.\EOS\space}%
\providecommand \EOS [0]{\spacefactor3000\relax}%
\providecommand \BibitemShut  [1]{\csname bibitem#1\endcsname}%
\let\auto@bib@innerbib\@empty
%</preamble>
\bibitem [{\citenamefont {Bernevig}\ \emph {et~al.}(2006)\citenamefont
  {Bernevig}, \citenamefont {Hughes},\ and\ \citenamefont
  {Zhang}}]{Bernevig2006}%
  \BibitemOpen
  \bibfield  {author} {\bibinfo {author} {\bibfnamefont {B.~A.}\ \bibnamefont
  {Bernevig}}, \bibinfo {author} {\bibfnamefont {T.~L.}\ \bibnamefont
  {Hughes}}, \ and\ \bibinfo {author} {\bibfnamefont {S.-C.}\ \bibnamefont
  {Zhang}},\ }\href {\doibase 10.1126/science.1133734} {\bibfield  {journal}
  {\bibinfo  {journal} {Science}\ }\textbf {\bibinfo {volume} {314}},\ \bibinfo
  {pages} {1757} (\bibinfo {year} {2006})}\BibitemShut {NoStop}%
\bibitem [{\citenamefont {K\"onig}\ \emph {et~al.}(2007)\citenamefont
  {K\"onig}, \citenamefont {Wiedmann}, \citenamefont {Brune}, \citenamefont
  {Roth}, \citenamefont {Buhmann}, \citenamefont {Molenkamp}, \citenamefont
  {Qi},\ and\ \citenamefont {Zhang}}]{Konig2007}%
  \BibitemOpen
  \bibfield  {author} {\bibinfo {author} {\bibfnamefont {M.}~\bibnamefont
  {K\"onig}}, \bibinfo {author} {\bibfnamefont {S.}~\bibnamefont {Wiedmann}},
  \bibinfo {author} {\bibfnamefont {C.}~\bibnamefont {Brune}}, \bibinfo
  {author} {\bibfnamefont {A.}~\bibnamefont {Roth}}, \bibinfo {author}
  {\bibfnamefont {H.}~\bibnamefont {Buhmann}}, \bibinfo {author} {\bibfnamefont
  {L.~W.}\ \bibnamefont {Molenkamp}}, \bibinfo {author} {\bibfnamefont {X.-L.}\
  \bibnamefont {Qi}}, \ and\ \bibinfo {author} {\bibfnamefont {S.-C.}\
  \bibnamefont {Zhang}},\ }\href {\doibase 10.1126/science.1148047} {\bibfield
  {journal} {\bibinfo  {journal} {Science}\ }\textbf {\bibinfo {volume}
  {318}},\ \bibinfo {pages} {766} (\bibinfo {year} {2007})}\BibitemShut
  {NoStop}%
\bibitem [{\citenamefont {Moore}(2010)}]{Moore2010}%
  \BibitemOpen
  \bibfield  {author} {\bibinfo {author} {\bibfnamefont {J.~E.}\ \bibnamefont
  {Moore}},\ }\href {\doibase 10.1038/nature08916} {\bibfield  {journal}
  {\bibinfo  {journal} {Nature}\ }\textbf {\bibinfo {volume} {464}},\ \bibinfo
  {pages} {194} (\bibinfo {year} {2010})}\BibitemShut {NoStop}%
\bibitem [{\citenamefont {Hasan}\ and\ \citenamefont {Kane}(2010)}]{Hasan2010}%
  \BibitemOpen
  \bibfield  {author} {\bibinfo {author} {\bibfnamefont {M.}~\bibnamefont
  {Hasan}}\ and\ \bibinfo {author} {\bibfnamefont {C.}~\bibnamefont {Kane}},\
  }\href {\doibase 10.1103/RevModPhys.82.3045} {\bibfield  {journal} {\bibinfo
  {journal} {Reviews of Modern Physics}\ }\textbf {\bibinfo {volume} {82}},\
  \bibinfo {pages} {3045} (\bibinfo {year} {2010})},\ \Eprint
  {http://arxiv.org/abs/1002.3895} {1002.3895} \BibitemShut {NoStop}%
\bibitem [{\citenamefont {Xiao}\ \emph {et~al.}(2010)\citenamefont {Xiao},
  \citenamefont {Yao}, \citenamefont {Feng}, \citenamefont {Wen}, \citenamefont
  {Zhu}, \citenamefont {Chen}, \citenamefont {Stocks},\ and\ \citenamefont
  {Zhang}}]{Xiao2010b}%
  \BibitemOpen
  \bibfield  {author} {\bibinfo {author} {\bibfnamefont {D.}~\bibnamefont
  {Xiao}}, \bibinfo {author} {\bibfnamefont {Y.}~\bibnamefont {Yao}}, \bibinfo
  {author} {\bibfnamefont {W.}~\bibnamefont {Feng}}, \bibinfo {author}
  {\bibfnamefont {J.}~\bibnamefont {Wen}}, \bibinfo {author} {\bibfnamefont
  {W.}~\bibnamefont {Zhu}}, \bibinfo {author} {\bibfnamefont {X.-Q.}\
  \bibnamefont {Chen}}, \bibinfo {author} {\bibfnamefont {G.}~\bibnamefont
  {Stocks}}, \ and\ \bibinfo {author} {\bibfnamefont {Z.}~\bibnamefont
  {Zhang}},\ }\href {\doibase 10.1103/PhysRevLett.105.096404} {\bibfield
  {journal} {\bibinfo  {journal} {Physical Review Letters}\ }\textbf {\bibinfo
  {volume} {105}},\ \bibinfo {pages} {96404} (\bibinfo {year}
  {2010})}\BibitemShut {NoStop}%
\bibitem [{\citenamefont {Hsieh}\ \emph {et~al.}(2009)\citenamefont {Hsieh},
  \citenamefont {Xia}, \citenamefont {Wray}, \citenamefont {Qian},
  \citenamefont {Pal}, \citenamefont {Dil}, \citenamefont {Osterwalder},
  \citenamefont {Meier}, \citenamefont {Bihlmayer}, \citenamefont {Kane},
  \citenamefont {Hor}, \citenamefont {Cava},\ and\ \citenamefont
  {Hasan}}]{Hsieh2009c}%
  \BibitemOpen
  \bibfield  {author} {\bibinfo {author} {\bibfnamefont {D.}~\bibnamefont
  {Hsieh}}, \bibinfo {author} {\bibfnamefont {Y.}~\bibnamefont {Xia}}, \bibinfo
  {author} {\bibfnamefont {L.}~\bibnamefont {Wray}}, \bibinfo {author}
  {\bibfnamefont {D.}~\bibnamefont {Qian}}, \bibinfo {author} {\bibfnamefont
  {A.}~\bibnamefont {Pal}}, \bibinfo {author} {\bibfnamefont {J.~H.}\
  \bibnamefont {Dil}}, \bibinfo {author} {\bibfnamefont {J.}~\bibnamefont
  {Osterwalder}}, \bibinfo {author} {\bibfnamefont {F.}~\bibnamefont {Meier}},
  \bibinfo {author} {\bibfnamefont {G.}~\bibnamefont {Bihlmayer}}, \bibinfo
  {author} {\bibfnamefont {C.~L.}\ \bibnamefont {Kane}}, \bibinfo {author}
  {\bibfnamefont {Y.~S.}\ \bibnamefont {Hor}}, \bibinfo {author} {\bibfnamefont
  {R.~J.}\ \bibnamefont {Cava}}, \ and\ \bibinfo {author} {\bibfnamefont
  {M.~Z.}\ \bibnamefont {Hasan}},\ }\href {\doibase 10.1126/science.1167733}
  {\bibfield  {journal} {\bibinfo  {journal} {Science}\ }\textbf {\bibinfo
  {volume} {323}},\ \bibinfo {pages} {919} (\bibinfo {year}
  {2009})}\BibitemShut {NoStop}%
\bibitem [{\citenamefont {Zhang}\ \emph {et~al.}(2009)\citenamefont {Zhang},
  \citenamefont {Liu}, \citenamefont {Qi}, \citenamefont {Deng}, \citenamefont
  {Dai}, \citenamefont {Zhang},\ and\ \citenamefont {Fang}}]{Zhang2009a}%
  \BibitemOpen
  \bibfield  {author} {\bibinfo {author} {\bibfnamefont {H.-J.}\ \bibnamefont
  {Zhang}}, \bibinfo {author} {\bibfnamefont {C.-X.}\ \bibnamefont {Liu}},
  \bibinfo {author} {\bibfnamefont {X.-L.}\ \bibnamefont {Qi}}, \bibinfo
  {author} {\bibfnamefont {X.-Y.}\ \bibnamefont {Deng}}, \bibinfo {author}
  {\bibfnamefont {X.}~\bibnamefont {Dai}}, \bibinfo {author} {\bibfnamefont
  {S.-C.}\ \bibnamefont {Zhang}}, \ and\ \bibinfo {author} {\bibfnamefont
  {Z.}~\bibnamefont {Fang}},\ }\href {\doibase 10.1103/PhysRevB.80.085307}
  {\bibfield  {journal} {\bibinfo  {journal} {Physical Review B}\ }\textbf
  {\bibinfo {volume} {80}},\ \bibinfo {pages} {1} (\bibinfo {year}
  {2009})}\BibitemShut {NoStop}%
\bibitem [{\citenamefont {Mong}\ \emph {et~al.}(2010)\citenamefont {Mong},
  \citenamefont {Essin},\ and\ \citenamefont {Moore}}]{Mong2010}%
  \BibitemOpen
  \bibfield  {author} {\bibinfo {author} {\bibfnamefont {R.~S.~K.}\
  \bibnamefont {Mong}}, \bibinfo {author} {\bibfnamefont {A.~M.}\ \bibnamefont
  {Essin}}, \ and\ \bibinfo {author} {\bibfnamefont {J.~E.}\ \bibnamefont
  {Moore}},\ }\href {\doibase 10.1103/PhysRevB.81.245209} {\bibfield  {journal}
  {\bibinfo  {journal} {Physical Review B}\ }\textbf {\bibinfo {volume} {81}},\
  \bibinfo {pages} {245209} (\bibinfo {year} {2010})}\BibitemShut {NoStop}%
\bibitem [{\citenamefont {Chadov}\ \emph {et~al.}(2010)\citenamefont {Chadov},
  \citenamefont {Qi}, \citenamefont {K\"{u}bler}, \citenamefont {Fecher},
  \citenamefont {Felser},\ and\ \citenamefont {Zhang}}]{Chadov2010}%
  \BibitemOpen
  \bibfield  {author} {\bibinfo {author} {\bibfnamefont {S.}~\bibnamefont
  {Chadov}}, \bibinfo {author} {\bibfnamefont {X.}~\bibnamefont {Qi}}, \bibinfo
  {author} {\bibfnamefont {J.}~\bibnamefont {K\"{u}bler}}, \bibinfo {author}
  {\bibfnamefont {G.~H.}\ \bibnamefont {Fecher}}, \bibinfo {author}
  {\bibfnamefont {C.}~\bibnamefont {Felser}}, \ and\ \bibinfo {author}
  {\bibfnamefont {S.~C.}\ \bibnamefont {Zhang}},\ }\href {\doibase
  10.1038/nmat2770} {\bibfield  {journal} {\bibinfo  {journal} {Nature
  Materials}\ }\textbf {\bibinfo {volume} {9}},\ \bibinfo {pages} {541}
  (\bibinfo {year} {2010})}\BibitemShut {NoStop}%
\bibitem [{\citenamefont {Al-Sawai}\ \emph {et~al.}(2010)\citenamefont
  {Al-Sawai}, \citenamefont {Lin}, \citenamefont {Markiewicz}, \citenamefont
  {Wray}, \citenamefont {Xia}, \citenamefont {Xu}, \citenamefont {Hasan},\ and\
  \citenamefont {Bansil}}]{Al-Sawai2010}%
  \BibitemOpen
  \bibfield  {author} {\bibinfo {author} {\bibfnamefont {W.}~\bibnamefont
  {Al-Sawai}}, \bibinfo {author} {\bibfnamefont {H.}~\bibnamefont {Lin}},
  \bibinfo {author} {\bibfnamefont {R.~S.}\ \bibnamefont {Markiewicz}},
  \bibinfo {author} {\bibfnamefont {L.~A.}\ \bibnamefont {Wray}}, \bibinfo
  {author} {\bibfnamefont {Y.}~\bibnamefont {Xia}}, \bibinfo {author}
  {\bibfnamefont {S.-Y.}\ \bibnamefont {Xu}}, \bibinfo {author} {\bibfnamefont
  {M.~Z.}\ \bibnamefont {Hasan}}, \ and\ \bibinfo {author} {\bibfnamefont
  {A.}~\bibnamefont {Bansil}},\ }\href {\doibase 10.1103/PhysRevB.82.125208}
  {\bibfield  {journal} {\bibinfo  {journal} {Physical Review B}\ }\textbf
  {\bibinfo {volume} {82}},\ \bibinfo {pages} {125208} (\bibinfo {year}
  {2010})}\BibitemShut {NoStop}%
\bibitem [{\citenamefont {Li}\ \emph {et~al.}(2011)\citenamefont {Li},
  \citenamefont {Lian},\ and\ \citenamefont {Jiang}}]{Li2011}%
  \BibitemOpen
  \bibfield  {author} {\bibinfo {author} {\bibfnamefont {C.}~\bibnamefont
  {Li}}, \bibinfo {author} {\bibfnamefont {J.}~\bibnamefont {Lian}}, \ and\
  \bibinfo {author} {\bibfnamefont {Q.}~\bibnamefont {Jiang}},\ }\href
  {\doibase 10.1103/PhysRevB.83.235125} {\bibfield  {journal} {\bibinfo
  {journal} {Physical Review B}\ }\textbf {\bibinfo {volume} {83}},\ \bibinfo
  {pages} {235125} (\bibinfo {year} {2011})}\BibitemShut {NoStop}%
\bibitem [{\citenamefont {Canfield}\ \emph {et~al.}(1991)\citenamefont
  {Canfield}, \citenamefont {Thompson}, \citenamefont {Beyermann},
  \citenamefont {Lacerda}, \citenamefont {Hundley}, \citenamefont {Peterson},
  \citenamefont {Fisk},\ and\ \citenamefont {Ott}}]{Canfield1991}%
  \BibitemOpen
  \bibfield  {author} {\bibinfo {author} {\bibfnamefont {P.~C.}\ \bibnamefont
  {Canfield}}, \bibinfo {author} {\bibfnamefont {J.~D.}\ \bibnamefont
  {Thompson}}, \bibinfo {author} {\bibfnamefont {W.~P.}\ \bibnamefont
  {Beyermann}}, \bibinfo {author} {\bibfnamefont {A.}~\bibnamefont {Lacerda}},
  \bibinfo {author} {\bibfnamefont {M.~F.}\ \bibnamefont {Hundley}}, \bibinfo
  {author} {\bibfnamefont {E.}~\bibnamefont {Peterson}}, \bibinfo {author}
  {\bibfnamefont {Z.}~\bibnamefont {Fisk}}, \ and\ \bibinfo {author}
  {\bibfnamefont {H.~R.}\ \bibnamefont {Ott}},\ }\href {\doibase
  10.1063/1.350141} {\bibfield  {journal} {\bibinfo  {journal} {Journal of
  Applied Physics}\ }\textbf {\bibinfo {volume} {70}},\ \bibinfo {pages} {5800}
  (\bibinfo {year} {1991})}\BibitemShut {NoStop}%
\bibitem [{\citenamefont {Fisk}\ \emph {et~al.}(1991)\citenamefont {Fisk},
  \citenamefont {Canfield}, \citenamefont {Beyermann}, \citenamefont
  {Thompson}, \citenamefont {Hundley}, \citenamefont {Ott}, \citenamefont
  {Felder}, \citenamefont {Maple}, \citenamefont {{Lopez de la Torre}},
  \citenamefont {Visani},\ and\ \citenamefont {Seaman}}]{Fisk1991}%
  \BibitemOpen
  \bibfield  {author} {\bibinfo {author} {\bibfnamefont {Z.}~\bibnamefont
  {Fisk}}, \bibinfo {author} {\bibfnamefont {P.}~\bibnamefont {Canfield}},
  \bibinfo {author} {\bibfnamefont {W.}~\bibnamefont {Beyermann}}, \bibinfo
  {author} {\bibfnamefont {J.}~\bibnamefont {Thompson}}, \bibinfo {author}
  {\bibfnamefont {M.}~\bibnamefont {Hundley}}, \bibinfo {author} {\bibfnamefont
  {H.}~\bibnamefont {Ott}}, \bibinfo {author} {\bibfnamefont {E.}~\bibnamefont
  {Felder}}, \bibinfo {author} {\bibfnamefont {M.}~\bibnamefont {Maple}},
  \bibinfo {author} {\bibfnamefont {M.}~\bibnamefont {{Lopez de la Torre}}},
  \bibinfo {author} {\bibfnamefont {P.}~\bibnamefont {Visani}}, \ and\ \bibinfo
  {author} {\bibfnamefont {C.}~\bibnamefont {Seaman}},\ }\href {\doibase
  10.1103/PhysRevLett.67.3310} {\bibfield  {journal} {\bibinfo  {journal}
  {Physical Review Letters}\ }\textbf {\bibinfo {volume} {67}},\ \bibinfo
  {pages} {3310} (\bibinfo {year} {1991})}\BibitemShut {NoStop}%
\bibitem [{\citenamefont {Goll}\ \emph {et~al.}(2002)\citenamefont {Goll},
  \citenamefont {Hagel}, \citenamefont {L\"{o}hneysen}, \citenamefont
  {Pietrus}, \citenamefont {Wanka}, \citenamefont {Wosnitza}, \citenamefont
  {Zwicknagl}, \citenamefont {Yoshino}, \citenamefont {Takabatake},\ and\
  \citenamefont {Jansen}}]{Goll2002}%
  \BibitemOpen
  \bibfield  {author} {\bibinfo {author} {\bibfnamefont {G.}~\bibnamefont
  {Goll}}, \bibinfo {author} {\bibfnamefont {J.}~\bibnamefont {Hagel}},
  \bibinfo {author} {\bibfnamefont {H.~v.}\ \bibnamefont {L\"{o}hneysen}},
  \bibinfo {author} {\bibfnamefont {T.}~\bibnamefont {Pietrus}}, \bibinfo
  {author} {\bibfnamefont {S.}~\bibnamefont {Wanka}}, \bibinfo {author}
  {\bibfnamefont {J.}~\bibnamefont {Wosnitza}}, \bibinfo {author}
  {\bibfnamefont {G.}~\bibnamefont {Zwicknagl}}, \bibinfo {author}
  {\bibfnamefont {T.}~\bibnamefont {Yoshino}}, \bibinfo {author} {\bibfnamefont
  {T.}~\bibnamefont {Takabatake}}, \ and\ \bibinfo {author} {\bibfnamefont
  {A.~G.~M.}\ \bibnamefont {Jansen}},\ }\href {\doibase
  10.1209/epl/i2002-00566-9} {\bibfield  {journal} {\bibinfo  {journal}
  {Europhysics Letters}\ }\textbf {\bibinfo {volume} {57}},\ \bibinfo {pages}
  {233} (\bibinfo {year} {2002})}\BibitemShut {NoStop}%
\bibitem [{\citenamefont {Tafti}\ \emph {et~al.}(2013)\citenamefont {Tafti},
  \citenamefont {Fujii}, \citenamefont {Juneau-Fecteau}, \citenamefont
  {{Ren\'{e} de Cotret}}, \citenamefont {Doiron-Leyraud}, \citenamefont
  {Asamitsu},\ and\ \citenamefont {Taillefer}}]{Tafti2013}%
  \BibitemOpen
  \bibfield  {author} {\bibinfo {author} {\bibfnamefont {F.~F.}\ \bibnamefont
  {Tafti}}, \bibinfo {author} {\bibfnamefont {T.}~\bibnamefont {Fujii}},
  \bibinfo {author} {\bibfnamefont {A.}~\bibnamefont {Juneau-Fecteau}},
  \bibinfo {author} {\bibfnamefont {S.}~\bibnamefont {{Ren\'{e} de Cotret}}},
  \bibinfo {author} {\bibfnamefont {N.}~\bibnamefont {Doiron-Leyraud}},
  \bibinfo {author} {\bibfnamefont {A.}~\bibnamefont {Asamitsu}}, \ and\
  \bibinfo {author} {\bibfnamefont {L.}~\bibnamefont {Taillefer}},\ }\href
  {\doibase 10.1103/PhysRevB.87.184504} {\bibfield  {journal} {\bibinfo
  {journal} {Physical Review B}\ }\textbf {\bibinfo {volume} {87}},\ \bibinfo
  {pages} {184504} (\bibinfo {year} {2013})}\BibitemShut {NoStop}%
\bibitem [{\citenamefont {Butch}\ \emph {et~al.}(2011)\citenamefont {Butch},
  \citenamefont {Syers}, \citenamefont {Kirshenbaum}, \citenamefont {Hope},\
  and\ \citenamefont {Paglione}}]{Butch2011}%
  \BibitemOpen
  \bibfield  {author} {\bibinfo {author} {\bibfnamefont {N.}~\bibnamefont
  {Butch}}, \bibinfo {author} {\bibfnamefont {P.}~\bibnamefont {Syers}},
  \bibinfo {author} {\bibfnamefont {K.}~\bibnamefont {Kirshenbaum}}, \bibinfo
  {author} {\bibfnamefont {A.}~\bibnamefont {Hope}}, \ and\ \bibinfo {author}
  {\bibfnamefont {J.}~\bibnamefont {Paglione}},\ }\href {\doibase
  10.1103/PhysRevB.84.220504} {\bibfield  {journal} {\bibinfo  {journal}
  {Physical Review B}\ }\textbf {\bibinfo {volume} {84}},\ \bibinfo {pages} {1}
  (\bibinfo {year} {2011})}\BibitemShut {NoStop}%
\bibitem [{\citenamefont {Bay}\ \emph {et~al.}(2012)\citenamefont {Bay},
  \citenamefont {Naka}, \citenamefont {Huang},\ and\ \citenamefont
  {de~Visser}}]{Bay2012}%
  \BibitemOpen
  \bibfield  {author} {\bibinfo {author} {\bibfnamefont {T.~V.}\ \bibnamefont
  {Bay}}, \bibinfo {author} {\bibfnamefont {T.}~\bibnamefont {Naka}}, \bibinfo
  {author} {\bibfnamefont {Y.~K.}\ \bibnamefont {Huang}}, \ and\ \bibinfo
  {author} {\bibfnamefont {A.}~\bibnamefont {de~Visser}},\ }\href {\doibase
  10.1103/PhysRevB.86.064515} {\bibfield  {journal} {\bibinfo  {journal}
  {Physical Review B}\ }\textbf {\bibinfo {volume} {86}},\ \bibinfo {pages}
  {064515} (\bibinfo {year} {2012})}\BibitemShut {NoStop}%
\bibitem [{\citenamefont {Bay}\ \emph {et~al.}(2014)\citenamefont {Bay},
  \citenamefont {Jackson}, \citenamefont {Paulsen}, \citenamefont {Baines},
  \citenamefont {Amato}, \citenamefont {Orvis}, \citenamefont {Aronson},
  \citenamefont {Huang},\ and\ \citenamefont {de~Visser}}]{Bay2014}%
  \BibitemOpen
  \bibfield  {author} {\bibinfo {author} {\bibfnamefont {T.}~\bibnamefont
  {Bay}}, \bibinfo {author} {\bibfnamefont {M.}~\bibnamefont {Jackson}},
  \bibinfo {author} {\bibfnamefont {C.}~\bibnamefont {Paulsen}}, \bibinfo
  {author} {\bibfnamefont {C.}~\bibnamefont {Baines}}, \bibinfo {author}
  {\bibfnamefont {A.}~\bibnamefont {Amato}}, \bibinfo {author} {\bibfnamefont
  {T.}~\bibnamefont {Orvis}}, \bibinfo {author} {\bibfnamefont
  {M.}~\bibnamefont {Aronson}}, \bibinfo {author} {\bibfnamefont
  {Y.}~\bibnamefont {Huang}}, \ and\ \bibinfo {author} {\bibfnamefont
  {A.}~\bibnamefont {de~Visser}},\ }\href {\doibase 10.1016/j.ssc.2013.12.010}
  {\bibfield  {journal} {\bibinfo  {journal} {Solid State Communications}\
  }\textbf {\bibinfo {volume} {183}},\ \bibinfo {pages} {13} (\bibinfo {year}
  {2014})}\BibitemShut {NoStop}%
\bibitem [{\citenamefont {Kozlova}\ \emph {et~al.}(2005)\citenamefont
  {Kozlova}, \citenamefont {Hagel}, \citenamefont {Doerr}, \citenamefont
  {Wosnitza}, \citenamefont {Eckert}, \citenamefont {M\"{u}ller}, \citenamefont
  {Schultz}, \citenamefont {Opahle}, \citenamefont {Elgazzar}, \citenamefont
  {Richter}, \citenamefont {Goll}, \citenamefont {L\"{o}hneysen}, \citenamefont
  {Zwicknagl}, \citenamefont {Yoshino},\ and\ \citenamefont
  {Takabatake}}]{Kozlova2005}%
  \BibitemOpen
  \bibfield  {author} {\bibinfo {author} {\bibfnamefont {N.}~\bibnamefont
  {Kozlova}}, \bibinfo {author} {\bibfnamefont {J.}~\bibnamefont {Hagel}},
  \bibinfo {author} {\bibfnamefont {M.}~\bibnamefont {Doerr}}, \bibinfo
  {author} {\bibfnamefont {J.}~\bibnamefont {Wosnitza}}, \bibinfo {author}
  {\bibfnamefont {D.}~\bibnamefont {Eckert}}, \bibinfo {author} {\bibfnamefont
  {K.-H.}\ \bibnamefont {M\"{u}ller}}, \bibinfo {author} {\bibfnamefont
  {L.}~\bibnamefont {Schultz}}, \bibinfo {author} {\bibfnamefont
  {I.}~\bibnamefont {Opahle}}, \bibinfo {author} {\bibfnamefont
  {S.}~\bibnamefont {Elgazzar}}, \bibinfo {author} {\bibfnamefont
  {M.}~\bibnamefont {Richter}}, \bibinfo {author} {\bibfnamefont
  {G.}~\bibnamefont {Goll}}, \bibinfo {author} {\bibfnamefont {H.~v.}\
  \bibnamefont {L\"{o}hneysen}}, \bibinfo {author} {\bibfnamefont
  {G.}~\bibnamefont {Zwicknagl}}, \bibinfo {author} {\bibfnamefont
  {T.}~\bibnamefont {Yoshino}}, \ and\ \bibinfo {author} {\bibfnamefont
  {T.}~\bibnamefont {Takabatake}},\ }\href {\doibase
  10.1103/PhysRevLett.95.086403} {\bibfield  {journal} {\bibinfo  {journal}
  {Physical Review Letters}\ }\textbf {\bibinfo {volume} {95}},\ \bibinfo
  {pages} {086403} (\bibinfo {year} {2005})}\BibitemShut {NoStop}%
\bibitem [{\citenamefont {Gofryk}\ \emph {et~al.}(2007)\citenamefont {Gofryk},
  \citenamefont {Kaczorowski}, \citenamefont {Plackowski}, \citenamefont
  {Mucha}, \citenamefont {Leithe-Jasper}, \citenamefont {Schnelle},\ and\
  \citenamefont {Grin}}]{Gofryk2007}%
  \BibitemOpen
  \bibfield  {author} {\bibinfo {author} {\bibfnamefont {K.}~\bibnamefont
  {Gofryk}}, \bibinfo {author} {\bibfnamefont {D.}~\bibnamefont {Kaczorowski}},
  \bibinfo {author} {\bibfnamefont {T.}~\bibnamefont {Plackowski}}, \bibinfo
  {author} {\bibfnamefont {J.}~\bibnamefont {Mucha}}, \bibinfo {author}
  {\bibfnamefont {A.}~\bibnamefont {Leithe-Jasper}}, \bibinfo {author}
  {\bibfnamefont {W.}~\bibnamefont {Schnelle}}, \ and\ \bibinfo {author}
  {\bibfnamefont {Y.}~\bibnamefont {Grin}},\ }\href {\doibase
  10.1103/PhysRevB.75.224426} {\bibfield  {journal} {\bibinfo  {journal}
  {Physical Review B}\ }\textbf {\bibinfo {volume} {75}},\ \bibinfo {pages}
  {224426} (\bibinfo {year} {2007})}\BibitemShut {NoStop}%
\bibitem [{\citenamefont {Xu}\ \emph {et~al.}(2014)\citenamefont {Xu},
  \citenamefont {Wang}, \citenamefont {Zhang}, \citenamefont {Du},
  \citenamefont {Liu}, \citenamefont {Wang}, \citenamefont {Wu}, \citenamefont
  {Liu},\ and\ \citenamefont {Zhang}}]{Xu2014}%
  \BibitemOpen
  \bibfield  {author} {\bibinfo {author} {\bibfnamefont {G.}~\bibnamefont
  {Xu}}, \bibinfo {author} {\bibfnamefont {W.}~\bibnamefont {Wang}}, \bibinfo
  {author} {\bibfnamefont {X.}~\bibnamefont {Zhang}}, \bibinfo {author}
  {\bibfnamefont {Y.}~\bibnamefont {Du}}, \bibinfo {author} {\bibfnamefont
  {E.}~\bibnamefont {Liu}}, \bibinfo {author} {\bibfnamefont {S.}~\bibnamefont
  {Wang}}, \bibinfo {author} {\bibfnamefont {G.}~\bibnamefont {Wu}}, \bibinfo
  {author} {\bibfnamefont {Z.}~\bibnamefont {Liu}}, \ and\ \bibinfo {author}
  {\bibfnamefont {X.~X.}\ \bibnamefont {Zhang}},\ }\href {\doibase
  10.1038/srep05709} {\bibfield  {journal} {\bibinfo  {journal} {Scientific
  Reports}\ }\textbf {\bibinfo {volume} {4}},\ \bibinfo {pages} {5709}
  (\bibinfo {year} {2014})}\BibitemShut {NoStop}%
\bibitem [{\citenamefont {Liu}\ \emph {et~al.}(2011)\citenamefont {Liu},
  \citenamefont {Lee}, \citenamefont {Kondo}, \citenamefont {Mun},
  \citenamefont {Caudle}, \citenamefont {Harmon}, \citenamefont {Bud’ko},
  \citenamefont {Canfield},\ and\ \citenamefont {Kaminski}}]{Liu2011}%
  \BibitemOpen
  \bibfield  {author} {\bibinfo {author} {\bibfnamefont {C.}~\bibnamefont
  {Liu}}, \bibinfo {author} {\bibfnamefont {Y.}~\bibnamefont {Lee}}, \bibinfo
  {author} {\bibfnamefont {T.}~\bibnamefont {Kondo}}, \bibinfo {author}
  {\bibfnamefont {E.}~\bibnamefont {Mun}}, \bibinfo {author} {\bibfnamefont
  {M.}~\bibnamefont {Caudle}}, \bibinfo {author} {\bibfnamefont
  {B.}~\bibnamefont {Harmon}}, \bibinfo {author} {\bibfnamefont
  {S.}~\bibnamefont {Bud’ko}}, \bibinfo {author} {\bibfnamefont
  {P.}~\bibnamefont {Canfield}}, \ and\ \bibinfo {author} {\bibfnamefont
  {A.}~\bibnamefont {Kaminski}},\ }\href {\doibase 10.1103/PhysRevB.83.205133}
  {\bibfield  {journal} {\bibinfo  {journal} {Physical Review B}\ }\textbf
  {\bibinfo {volume} {83}},\ \bibinfo {pages} {205133} (\bibinfo {year}
  {2011})}\BibitemShut {NoStop}%
\bibitem [{\citenamefont {Kreyssig}\ \emph {et~al.}(2011)\citenamefont
  {Kreyssig}, \citenamefont {Kim}, \citenamefont {Kim}, \citenamefont {Pratt},
  \citenamefont {Sauerbrei}, \citenamefont {March}, \citenamefont {Tesdall},
  \citenamefont {Bud'ko}, \citenamefont {Canfield}, \citenamefont {McQueeney},\
  and\ \citenamefont {Goldman}}]{Kreyssig2011a}%
  \BibitemOpen
  \bibfield  {author} {\bibinfo {author} {\bibfnamefont {A.}~\bibnamefont
  {Kreyssig}}, \bibinfo {author} {\bibfnamefont {M.~G.}\ \bibnamefont {Kim}},
  \bibinfo {author} {\bibfnamefont {J.~D.}\ \bibnamefont {Kim}}, \bibinfo
  {author} {\bibfnamefont {D.~K.}\ \bibnamefont {Pratt}}, \bibinfo {author}
  {\bibfnamefont {S.~M.}\ \bibnamefont {Sauerbrei}}, \bibinfo {author}
  {\bibfnamefont {S.~D.}\ \bibnamefont {March}}, \bibinfo {author}
  {\bibfnamefont {G.~R.}\ \bibnamefont {Tesdall}}, \bibinfo {author}
  {\bibfnamefont {S.~L.}\ \bibnamefont {Bud'ko}}, \bibinfo {author}
  {\bibfnamefont {P.~C.}\ \bibnamefont {Canfield}}, \bibinfo {author}
  {\bibfnamefont {R.~J.}\ \bibnamefont {McQueeney}}, \ and\ \bibinfo {author}
  {\bibfnamefont {A.~I.}\ \bibnamefont {Goldman}},\ }\href {\doibase
  10.1103/PhysRevB.84.220408} {\bibfield  {journal} {\bibinfo  {journal}
  {Physical Review B}\ }\textbf {\bibinfo {volume} {84}},\ \bibinfo {pages}
  {220408} (\bibinfo {year} {2011})}\BibitemShut {NoStop}%
\bibitem [{\citenamefont {M\"{u}ller}\ \emph {et~al.}(2014)\citenamefont
  {M\"{u}ller}, \citenamefont {Lee-Hone}, \citenamefont {Lapointe},
  \citenamefont {Ryan}, \citenamefont {Pereg-Barnea}, \citenamefont {Bianchi},
  \citenamefont {Mozharivskyj},\ and\ \citenamefont {Flacau}}]{Muller2014}%
  \BibitemOpen
  \bibfield  {author} {\bibinfo {author} {\bibfnamefont {R.~A.}\ \bibnamefont
  {M\"{u}ller}}, \bibinfo {author} {\bibfnamefont {N.~R.}\ \bibnamefont
  {Lee-Hone}}, \bibinfo {author} {\bibfnamefont {L.}~\bibnamefont {Lapointe}},
  \bibinfo {author} {\bibfnamefont {D.~H.}\ \bibnamefont {Ryan}}, \bibinfo
  {author} {\bibfnamefont {T.}~\bibnamefont {Pereg-Barnea}}, \bibinfo {author}
  {\bibfnamefont {A.~D.}\ \bibnamefont {Bianchi}}, \bibinfo {author}
  {\bibfnamefont {Y.}~\bibnamefont {Mozharivskyj}}, \ and\ \bibinfo {author}
  {\bibfnamefont {R.}~\bibnamefont {Flacau}},\ }\href {\doibase
  10.1103/PhysRevB.90.041109} {\bibfield  {journal} {\bibinfo  {journal}
  {Physical Review B}\ }\textbf {\bibinfo {volume} {90}},\ \bibinfo {pages}
  {041109} (\bibinfo {year} {2014})}\BibitemShut {NoStop}%
\bibitem [{APE(2009)}]{APEX}%
  \BibitemOpen
  \href@noop {} {\emph {\bibinfo {title} {{APEX2 and SAINT, Version 7.68A}}}}\
  (\bibinfo  {publisher} {Bruker AXS Inc.},\ \bibinfo {address} {Madison, WI},\
  \bibinfo {year} {2009})\BibitemShut {NoStop}%
\bibitem [{\citenamefont {Sheldrick}(2008{\natexlab{a}})}]{Sheldrick2008a}%
  \BibitemOpen
  \bibfield  {author} {\bibinfo {author} {\bibfnamefont {G.~M.}\ \bibnamefont
  {Sheldrick}},\ }\href@noop {} {\emph {\bibinfo {title} {{SADABS, Version
  2008}}}}\ (\bibinfo  {publisher} {Bruker AXS Inc.},\ \bibinfo {address}
  {Madison, WI},\ \bibinfo {year} {2008})\BibitemShut {NoStop}%
\bibitem [{\citenamefont {Sheldrick}(2008{\natexlab{b}})}]{Sheldrick2008}%
  \BibitemOpen
  \bibfield  {author} {\bibinfo {author} {\bibfnamefont {G.~M.}\ \bibnamefont
  {Sheldrick}},\ }\href {\doibase 10.1107/S0108767307043930} {\bibfield
  {journal} {\bibinfo  {journal} {Acta crystallographica. Section A,
  Foundations of crystallography}\ }\textbf {\bibinfo {volume} {64}},\ \bibinfo
  {pages} {112} (\bibinfo {year} {2008}{\natexlab{b}})}\BibitemShut {NoStop}%
\bibitem [{\citenamefont {Wosnitza}\ \emph {et~al.}(2006)\citenamefont
  {Wosnitza}, \citenamefont {Goll}, \citenamefont {Bianchi}, \citenamefont
  {Bergk}, \citenamefont {Kozlova}, \citenamefont {Opahle}, \citenamefont
  {Elgazzar}, \citenamefont {Richter}, \citenamefont {Stockert}, \citenamefont
  {L\"{o}hneysen}, \citenamefont {Yoshino},\ and\ \citenamefont
  {Takabatake}}]{Wosnitza2006}%
  \BibitemOpen
  \bibfield  {author} {\bibinfo {author} {\bibfnamefont {J.}~\bibnamefont
  {Wosnitza}}, \bibinfo {author} {\bibfnamefont {G.}~\bibnamefont {Goll}},
  \bibinfo {author} {\bibfnamefont {A.~D.}\ \bibnamefont {Bianchi}}, \bibinfo
  {author} {\bibfnamefont {B.}~\bibnamefont {Bergk}}, \bibinfo {author}
  {\bibfnamefont {N.}~\bibnamefont {Kozlova}}, \bibinfo {author} {\bibfnamefont
  {I.}~\bibnamefont {Opahle}}, \bibinfo {author} {\bibfnamefont
  {S.}~\bibnamefont {Elgazzar}}, \bibinfo {author} {\bibfnamefont
  {M.}~\bibnamefont {Richter}}, \bibinfo {author} {\bibfnamefont
  {O.}~\bibnamefont {Stockert}}, \bibinfo {author} {\bibfnamefont {H.~v.}\
  \bibnamefont {L\"{o}hneysen}}, \bibinfo {author} {\bibfnamefont
  {T.}~\bibnamefont {Yoshino}}, \ and\ \bibinfo {author} {\bibfnamefont
  {T.}~\bibnamefont {Takabatake}},\ }\href {\doibase 10.1088/1367-2630/8/9/174}
  {\bibfield  {journal} {\bibinfo  {journal} {New Journal of Physics}\ }\textbf
  {\bibinfo {volume} {8}},\ \bibinfo {pages} {174} (\bibinfo {year}
  {2006})}\BibitemShut {NoStop}%
\bibitem [{\citenamefont {Morelli}\ \emph {et~al.}(1996)\citenamefont
  {Morelli}, \citenamefont {Canfield},\ and\ \citenamefont
  {Drymiotis}}]{Morelli1996}%
  \BibitemOpen
  \bibfield  {author} {\bibinfo {author} {\bibfnamefont {D.}~\bibnamefont
  {Morelli}}, \bibinfo {author} {\bibfnamefont {P.}~\bibnamefont {Canfield}}, \
  and\ \bibinfo {author} {\bibfnamefont {P.}~\bibnamefont {Drymiotis}},\ }\href
  {\doibase 10.1103/PhysRevB.53.12896} {\bibfield  {journal} {\bibinfo
  {journal} {Physical Review B}\ }\textbf {\bibinfo {volume} {53}},\ \bibinfo
  {pages} {12896} (\bibinfo {year} {1996})}\BibitemShut {NoStop}%
\bibitem [{\citenamefont {Fisher}\ and\ \citenamefont
  {Langer}(1968)}]{Fisher1968}%
  \BibitemOpen
  \bibfield  {author} {\bibinfo {author} {\bibfnamefont {M.}~\bibnamefont
  {Fisher}}\ and\ \bibinfo {author} {\bibfnamefont {J.}~\bibnamefont
  {Langer}},\ }\href {\doibase 10.1103/PhysRevLett.20.665} {\bibfield
  {journal} {\bibinfo  {journal} {Physical Review Letters}\ }\textbf {\bibinfo
  {volume} {20}},\ \bibinfo {pages} {665} (\bibinfo {year} {1968})}\BibitemShut
  {NoStop}%
\bibitem [{\citenamefont {Yildirim}\ \emph {et~al.}(1998)\citenamefont
  {Yildirim}, \citenamefont {Harris},\ and\ \citenamefont
  {Shender}}]{Yildirim1998}%
  \BibitemOpen
  \bibfield  {author} {\bibinfo {author} {\bibfnamefont {T.}~\bibnamefont
  {Yildirim}}, \bibinfo {author} {\bibfnamefont {A.}~\bibnamefont {Harris}}, \
  and\ \bibinfo {author} {\bibfnamefont {E.}~\bibnamefont {Shender}},\ }\href
  {\doibase 10.1103/PhysRevB.58.3144} {\bibfield  {journal} {\bibinfo
  {journal} {Physical Review B}\ }\textbf {\bibinfo {volume} {58}},\ \bibinfo
  {pages} {3144} (\bibinfo {year} {1998})}\BibitemShut {NoStop}%
\bibitem [{\citenamefont {Rossat-Mignod}(1987)}]{Rossat-Mignod1987}%
  \BibitemOpen
  \bibfield  {author} {\bibinfo {author} {\bibfnamefont {J.}~\bibnamefont
  {Rossat-Mignod}},\ }in\ \href {\doibase 10.1016/S0076-695X(08)60770-X} {\emph
  {\bibinfo {booktitle} {Methods in Experimental Physics}}}\ (\bibinfo
  {publisher} {Academic Press},\ \bibinfo {address} {New York},\ \bibinfo
  {year} {1987})\ Chap.~\bibinfo {chapter} {19}, pp.\ \bibinfo {pages}
  {69--157}\BibitemShut {NoStop}%
\bibitem [{\citenamefont {Rodriguez-Carvajal}(1993)}]{Rodriguez-carvajal1993}%
  \BibitemOpen
  \bibfield  {author} {\bibinfo {author} {\bibfnamefont {J.}~\bibnamefont
  {Rodriguez-Carvajal}},\ }\href@noop {} {\bibfield  {journal} {\bibinfo
  {journal} {Physica B}\ }\textbf {\bibinfo {volume} {192}},\ \bibinfo {pages}
  {55} (\bibinfo {year} {1993})}\BibitemShut {NoStop}%
\bibitem [{\citenamefont {Shirane}\ \emph {et~al.}(2002)\citenamefont
  {Shirane}, \citenamefont {Shapiro},\ and\ \citenamefont
  {Tranquada}}]{Shirane2002}%
  \BibitemOpen
  \bibfield  {author} {\bibinfo {author} {\bibfnamefont {G.}~\bibnamefont
  {Shirane}}, \bibinfo {author} {\bibfnamefont {S.~M.}\ \bibnamefont
  {Shapiro}}, \ and\ \bibinfo {author} {\bibfnamefont {J.~M.}\ \bibnamefont
  {Tranquada}},\ }\href@noop {} {\emph {\bibinfo {title} {{Neutron Scattering
  with a Triple Axis Spectrometer}}}}\ (\bibinfo  {publisher} {Cambridge
  University Press},\ \bibinfo {address} {Cambridge},\ \bibinfo {year}
  {2002})\BibitemShut {NoStop}%
\bibitem [{\citenamefont {Pelissetto}\ and\ \citenamefont
  {Vicari}(2002)}]{Pelissetto2002}%
  \BibitemOpen
  \bibfield  {author} {\bibinfo {author} {\bibfnamefont {A.}~\bibnamefont
  {Pelissetto}}\ and\ \bibinfo {author} {\bibfnamefont {E.}~\bibnamefont
  {Vicari}},\ }\href {\doibase 10.1016/S0370-1573(02)00219-3} {\bibfield
  {journal} {\bibinfo  {journal} {Physics Reports}\ }\textbf {\bibinfo {volume}
  {368}},\ \bibinfo {pages} {549} (\bibinfo {year} {2002})}\BibitemShut
  {NoStop}%
\bibitem [{\citenamefont {Goll}\ \emph {et~al.}(2007)\citenamefont {Goll},
  \citenamefont {Stockert}, \citenamefont {Prager}, \citenamefont {Yoshino},\
  and\ \citenamefont {Takabatake}}]{Goll2007}%
  \BibitemOpen
  \bibfield  {author} {\bibinfo {author} {\bibfnamefont {G.}~\bibnamefont
  {Goll}}, \bibinfo {author} {\bibfnamefont {O.}~\bibnamefont {Stockert}},
  \bibinfo {author} {\bibfnamefont {M.}~\bibnamefont {Prager}}, \bibinfo
  {author} {\bibfnamefont {T.}~\bibnamefont {Yoshino}}, \ and\ \bibinfo
  {author} {\bibfnamefont {T.}~\bibnamefont {Takabatake}},\ }\href {\doibase
  10.1016/j.jmmm.2006.10.1085} {\bibfield  {journal} {\bibinfo  {journal}
  {Journal of Magnetism and Magnetic Materials}\ }\textbf {\bibinfo {volume}
  {310}},\ \bibinfo {pages} {1773} (\bibinfo {year} {2007})}\BibitemShut
  {NoStop}%
\bibitem [{\citenamefont {Lea}\ \emph {et~al.}(1962)\citenamefont {Lea},
  \citenamefont {Leask},\ and\ \citenamefont {Wolf}}]{Lea1962}%
  \BibitemOpen
  \bibfield  {author} {\bibinfo {author} {\bibfnamefont {K.}~\bibnamefont
  {Lea}}, \bibinfo {author} {\bibfnamefont {M.}~\bibnamefont {Leask}}, \ and\
  \bibinfo {author} {\bibfnamefont {W.}~\bibnamefont {Wolf}},\ }\href {\doibase
  10.1016/0022-3697(62)90192-0} {\bibfield  {journal} {\bibinfo  {journal}
  {Journal of Physics and Chemistry of Solids}\ }\textbf {\bibinfo {volume}
  {23}},\ \bibinfo {pages} {1381} (\bibinfo {year} {1962})}\BibitemShut
  {NoStop}%
\bibitem [{\citenamefont {Ueland}\ \emph {et~al.}(2014)\citenamefont {Ueland},
  \citenamefont {Kreyssig}, \citenamefont {Proke\v{s}}, \citenamefont {Lynn},
  \citenamefont {Harriger}, \citenamefont {Pratt}, \citenamefont {Singh},
  \citenamefont {Heitmann}, \citenamefont {Sauerbrei}, \citenamefont
  {Saunders}, \citenamefont {Mun}, \citenamefont {Bud'ko}, \citenamefont
  {McQueeney}, \citenamefont {Canfield},\ and\ \citenamefont
  {Goldman}}]{Ueland2014}%
  \BibitemOpen
  \bibfield  {author} {\bibinfo {author} {\bibfnamefont {B.~G.}\ \bibnamefont
  {Ueland}}, \bibinfo {author} {\bibfnamefont {A.}~\bibnamefont {Kreyssig}},
  \bibinfo {author} {\bibfnamefont {K.}~\bibnamefont {Proke\v{s}}}, \bibinfo
  {author} {\bibfnamefont {J.~W.}\ \bibnamefont {Lynn}}, \bibinfo {author}
  {\bibfnamefont {L.~W.}\ \bibnamefont {Harriger}}, \bibinfo {author}
  {\bibfnamefont {D.~K.}\ \bibnamefont {Pratt}}, \bibinfo {author}
  {\bibfnamefont {D.~K.}\ \bibnamefont {Singh}}, \bibinfo {author}
  {\bibfnamefont {T.~W.}\ \bibnamefont {Heitmann}}, \bibinfo {author}
  {\bibfnamefont {S.}~\bibnamefont {Sauerbrei}}, \bibinfo {author}
  {\bibfnamefont {S.~M.}\ \bibnamefont {Saunders}}, \bibinfo {author}
  {\bibfnamefont {E.~D.}\ \bibnamefont {Mun}}, \bibinfo {author} {\bibfnamefont
  {S.~L.}\ \bibnamefont {Bud'ko}}, \bibinfo {author} {\bibfnamefont {R.~J.}\
  \bibnamefont {McQueeney}}, \bibinfo {author} {\bibfnamefont {P.~C.}\
  \bibnamefont {Canfield}}, \ and\ \bibinfo {author} {\bibfnamefont {A.~I.}\
  \bibnamefont {Goldman}},\ }\href {\doibase 10.1103/PhysRevB.89.180403}
  {\bibfield  {journal} {\bibinfo  {journal} {Physical Review B}\ }\textbf
  {\bibinfo {volume} {89}},\ \bibinfo {pages} {180403} (\bibinfo {year}
  {2014})}\BibitemShut {NoStop}%
\bibitem [{\citenamefont {Forster}\ \emph {et~al.}(1968)\citenamefont
  {Forster}, \citenamefont {Johnston},\ and\ \citenamefont
  {Wheeler}}]{Forster1968}%
  \BibitemOpen
  \bibfield  {author} {\bibinfo {author} {\bibfnamefont {R.}~\bibnamefont
  {Forster}}, \bibinfo {author} {\bibfnamefont {G.}~\bibnamefont {Johnston}}, \
  and\ \bibinfo {author} {\bibfnamefont {D.}~\bibnamefont {Wheeler}},\ }\href
  {\doibase 10.1016/0022-3697(68)90147-9} {\bibfield  {journal} {\bibinfo
  {journal} {Journal of Physics and Chemistry of Solids}\ }\textbf {\bibinfo
  {volume} {29}},\ \bibinfo {pages} {855} (\bibinfo {year} {1968})}\BibitemShut
  {NoStop}%
\bibitem [{\citenamefont {Halder}\ \emph {et~al.}(2011)\citenamefont {Halder},
  \citenamefont {Yusuf}, \citenamefont {Kumar}, \citenamefont {Nigam},\ and\
  \citenamefont {Keller}}]{Halder2011}%
  \BibitemOpen
  \bibfield  {author} {\bibinfo {author} {\bibfnamefont {M.}~\bibnamefont
  {Halder}}, \bibinfo {author} {\bibfnamefont {S.~M.}\ \bibnamefont {Yusuf}},
  \bibinfo {author} {\bibfnamefont {A.}~\bibnamefont {Kumar}}, \bibinfo
  {author} {\bibfnamefont {A.~K.}\ \bibnamefont {Nigam}}, \ and\ \bibinfo
  {author} {\bibfnamefont {L.}~\bibnamefont {Keller}},\ }\href {\doibase
  10.1103/PhysRevB.84.094435} {\bibfield  {journal} {\bibinfo  {journal}
  {Physical Review B}\ }\textbf {\bibinfo {volume} {84}},\ \bibinfo {pages}
  {094435} (\bibinfo {year} {2011})}\BibitemShut {NoStop}%
\end{thebibliography}%
